\documentclass[aps,pre,reprint,floatfix,longbibliography,superscriptaddress,amsmath,amssymb,compress]{revtex4-1}
\pdfoutput=1

\usepackage{hyperref}

\usepackage{graphicx}
\usepackage{color}

\usepackage{booktabs}

\usepackage{siunitx}

\usepackage[dvipsnames]{xcolor}

\AtBeginDocument{
  \heavyrulewidth=.08em
  \lightrulewidth=.06em
  \cmidrulewidth=.03em
  \belowrulesep=.65ex
  \belowbottomsep=0pt
  \aboverulesep=.4ex
  \abovetopsep=0pt
  \cmidrulesep=\doublerulesep
  \cmidrulekern=.5em
  \defaultaddspace=.5em
  }

\newcommand{\papertitle}{Delocalization of interacting directed polymers on a periodic substrate: Localization length and critical exponents from non-Hermitian spectra}

\newcommand{\subfiglabel}[1]{{\bfseries #1}}
\newcommand{\figref}[1]{Fig.~\ref{#1}}
\newcommand{\subfigref}[2]{\figref{#1}\subfiglabel{#2}}

\newcommand{\appref}[1]{Appendix~\ref{#1}}
\newcommand{\eqnref}[1]{Eq.~\eqref{#1}}

\newcommand{\tcrit}{\theta_\text{c}}
\newcommand{\sigmad}{\sigma_\text{d}}

\newcommand{\Psit}{\widetilde \Psi}
\newcommand{\ket}[1]{{| #1 \rangle}}
\newcommand{\bra}[1]{{\langle #1 |}}

\newcommand{\expect}[1]{{\langle #1 \rangle}}
\newcommand{\bs}[1]{{\boldsymbol #1}}
\newcommand{\emb}{\bs{\varepsilon}}
\newcommand{\rea}{\operatorname{Re}}
\newcommand{\ima}{\operatorname{Im}}
\newcommand{\Hhn}{\mathcal{H}(g)} 
\newcommand{\Hmb}{\mathbf{H}} 

\newcommand{\acktext}{\emph{Author contributions:} A.M. performed the theoretical analysis and drafted the manuscript. A.P. developed and implemented the numerical simulations and analyzed simulation data. J.P. designed and supervised the research, analyzed simulation data, and edited the
manuscript.
  
This project originated in discussions with David R. Nelson, Vincenzo Vitelli, and Anton Souslov during the \emph{Topological Metamaterials} Winter Conference at the Aspen Center for Physics (January 2017). We thank John Toner for useful conversations. Work was supported by the National Science Foundation under Grant No.~DMR--2145766 and by start-up funds from the University of Oregon.} 

\newcommand{\affiliationA}{Department of Physics, University of Oregon, Eugene, Oregon 97403}
\newcommand{\affiliationB}{Institute for Fundamental Science and Materials Science Institute, University of Oregon, Eugene, Oregon 97403, USA}

\begin{document}
\title{\papertitle}
\author{Abhijeet Melkani}
\thanks{These authors contributed equally to this work.}
\affiliation{\affiliationA}
\affiliation{\affiliationB}
\author{Alexander Patapoff}
\thanks{These authors contributed equally to this work.}
\affiliation{\affiliationA}
\author{Jayson Paulose}
\email{jpaulose@uoregon.edu}
\affiliation{\affiliationA}
\affiliation{\affiliationB}
\begin{abstract}

We study a classical model of thermally fluctuating polymers confined to two dimensions, experiencing a grooved periodic potential, and subject to pulling forces both along and transverse to the grooves. The equilibrium polymer conformations are described by a mapping to a quantum system with a non-Hermitian Hamiltonian and with fermionic statistics generated by noncrossing interactions among polymers. Using molecular dynamics simulations and analytical calculations, we identify a localized and a delocalized phase of the polymer conformations, separated by a delocalization transition which corresponds (in the quantum description) to the breakdown of a band insulator when driven by an imaginary vector potential. We calculate the average tilt of the many-body system, at arbitrary shear values and filling density of polymer chains, in terms of the complex-valued non-Hermitian band structure. We find the critical shear value, the localization length, and the critical exponent by which the shear modulus diverges in terms of the branch points (exceptional points) in the band structure at which the bandgap closes. We also investigate the combined effects of non-Hermitian delocalization and localization due to both periodicity and disorder, uncovering preliminary evidence that while disorder favours localization at high values, it encourages delocalization at lower values.

\end{abstract}

\maketitle

\section{Introduction}

Non-Hermitian operators~\cite{ashida2020nonhermitian, bender2007nonhermitian} have been extensively used to describe the dynamics of a variety of quantum~\cite{lau2018sensing, elganainy2018nature, malzard2015photonic} as well as classical systems~\cite{schindler2012symmetric, thevamaran2019accoustic, Sone2019active, rosa2020feedback}. They are ubiquitous in both exact and effective models of nature capturing gain/loss in open systems~\cite{elganainy2018nature}, dissipation~\cite{li2019observation}, probability fluxes~\cite{nelson1998population,tang2021stochastic}, sensitivity to boundary conditions~\cite{borgnia2020boundary,okuma2020skin, yao2018edge}, and various other phenomena excluded by assumptions of Hermiticity. They also enable the description of new kinds of phase transitions and topological classifications beyond the existing Hermitian framework for condensed matter~\cite{gong2018topological, kawabata2019topology, liu2019topological}. For example, when non-Hermitian systems are periodic in space, their excitations are described by complex-valued band structures~\cite{shen2018topological} which support uniquely non-Hermitian properties such as exceptional points (branch points)~\cite{heiss2012exceptional}. The physical implications of non-Hermitian band effects have been explored in a wide range of classical systems~\cite{ghatak2019nonhermitian, zhou2020metamaterial,Scheibner2020,Shmuel2020, ramezani2012PTphotonic}.

In systems of many bodies---such as spin waves, electrons, polymer chains, and vortex lines---generic thermodynamic phases may be distinguished by the localization properties of probability densities throughout the bulk~\cite{abanin2019localization}. Here, non-Hermitian terms quantify the capacity of external forces and fields to generate fluxes of probability and information which can drive the system out of a localized phase~\cite{hatano1996localizationprl,hatano1997vortexPinning,nelson1998population, hatano1998delocalization, amir2016networks}. Non-Hermitian delocalization has been extensively studied in models of thermally fluctuating lines under simultaneous tension and shear forces in the presence of randomly positioned columnar pinning sites~\cite{hatano1996localizationprl, hatano1997vortexPinning}, which serve as effective models for magnetic vortex lines penetrating through type-II superconductors with columnar defects~\cite{nelson1990fluxLiquid, nelson1993boson, hwa1993disorder}. The statistical mechanics of fluctuating lines in $d$ dimensions was mapped to the quantum-mechanical time evolution of bosons in $d-1$ dimensions, whose Hamiltonian becomes non-Hermitian when the lines are sheared in a direction transverse to the defect axes. The single-particle energy eigenstates, which are localized due to disorder, become delocalized when the strength of non-Hermitian terms exceed a threshold corresponding to a critical shear force~\cite{hatano1996localizationprl}. The delocalization of the bosonic eigenstates manifests as a tilt in the average conformations of the lines relative to the columnar defects. This non-Hermitian delocalization transition survives in the presence of interactions~\cite{kim2001interaction, affleck2004pinning, refael2006meissner, lankhorst2018dynamic,hamazaki2019manybody,Panda2020}.

\begin{figure}[t]
  \centering
  \includegraphics{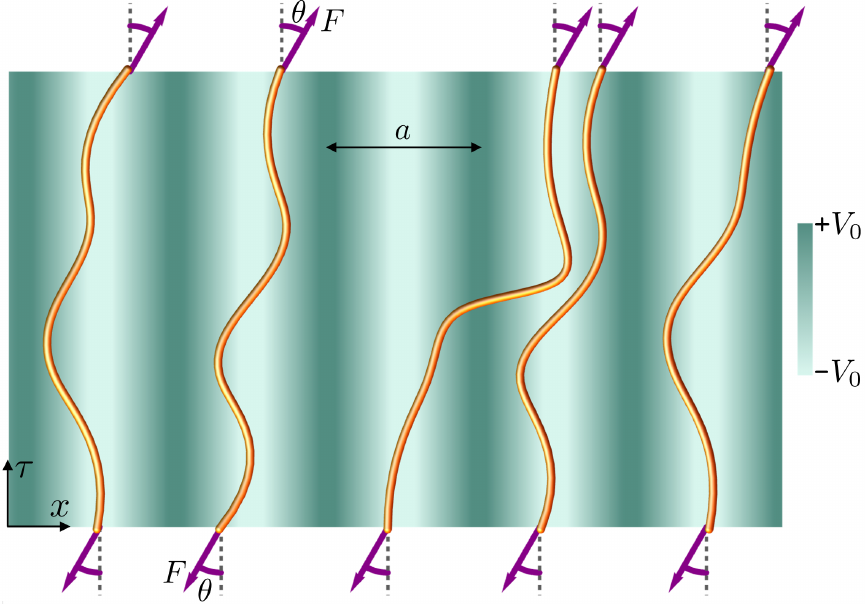}
  \caption{Schematic and description of our model system. Thermally fluctuating polymer chains (orange lines) with noncrossing constraints are subjected to a tension force $F$ (purple arrows) on a two-dimensional substrate potential of strength $V(x)$ per unit length (green background). The potential, of amplitude $V_0$, is periodic (with period $a$) along the $x$ direction and constant along the $\tau$ direction. The tension force, $F$, is applied to the ends of each chain at a specified force angle, $\theta$, with respect to the $\tau$-axis; the transverse component $F\theta$ of the tension is called the shear force. $F$ is assumed to be large enough such that the chains do not double back on themselves and self-interactions of a chain with itself are avoided.}
  \label{fig:intro}
\end{figure}

However, the simultaneous interplay of non-Hermitian drive, thermal fluctuations, interactions, and confinement due to spatial \emph{periodicity}, as opposed to \emph{disorder}, remains poorly understood. Prior works on non-Hermitian delocalization of magnetic vortices in periodic lattices of pinning sites operated in the tight-binding limit~\cite{fukui1998mott,hebert2011hatanoNelson, krachenbrink2021tilted}, thereby failing to capture dependencies on the form of the continuous potential~\cite{jayannavar1982disorder}. Field-theoretic studies, which in turn derived their effective action from the tight-binding limit via a Hubbard-Stratonovich transformation, uncovered new thermodynamic phases in the Hermitian setting~\cite{fisher1989boson} whose non-Hermitian counterparts have been investigated using a mean-field model in 2+1-dimensions (the upper critical dimension)~\cite{lehrer1998mott} but not in 1+1D. The form of the continuum equilibrium density profiles and their connection with spectral and topological features of non-Hermitian band structures was not elucidated in these prior studies.

In this work, we investigate non-Hermitian delocalization in a continuum statistical mechanical model with interactions, in which localization derives from a band-insulating state due to an underlying periodic potential. Specifically, we use classical molecular dynamics simulations and analytical calculations to study the effect of shear forces on a model of directed polymers confined to two dimensions and experiencing a smoothly varying periodic substrate potential (see \figref{fig:intro} for a schematic and a description of our model). Directed polymers are thermally fluctuating chains that are extended along a preferred direction by an external field, which prevents self-interactions within chains. Besides describing superconductor vortices~\cite{nelson1993boson,Bolle1999}, directed polymer models capture the statistical mechanics of semiflexible polymers embedded in liquid crystals~\cite{Dogic2004} and wandering steps on vicinal surfaces of crystals~\cite{Bartelt1990}. All these systems share the property that the extended constituents cannot cross each other in space. Directed polymers with noncrossing interactions can be exactly solved~\cite{gennes1968soluble,rocklin2012constraints}; in our model, noncrossing interactions combine with the periodic potential to generate a state in which individual polymers are localized to distinct grooves~\cite{pedro2019protection}. Upon increasing the shear strength, the polymers collectively undergo a delocalization transition at a threshold force angle beyond which their average equilibrium conformations are tilted and no longer align with the substrate.

We map our model system of polymer chains to a one-dimensional (1D) quantum Hamiltonian with non-Hermitian drive caused by an imaginary vector-potential term. We find the average tilt of the many-body system, at arbitrary shear values and filling density of polymer chains, in terms of the complex non-Hermitian band structure [\eqnref{eq:tilt-complex}] and show that the commensurate system undergoes a sharp transition in the thermodynamic limit. The delocalization transition corresponds to a gap closure in the complex non-Hermitian band structure associated with the substrate potential in the presence of shear forces. The exact value of the critical force angle at which the polymers delocalize is found in terms of the position of the branch point (exceptional point) in the spectrum [\eqnref{eq:critical_theory}] while the critical exponent by which the shear modulus diverges is determined by the order of the branch point and is universal for all periodic potentials [\eqnref{eq:phi-summary}]. The theoretical prediction of the critical force angle quantitatively agrees with the transition observed in our simulations.

Our theoretical predictions rely upon a gauge transformation which maps the non-Hermitian quantum system to a Hermitian system, albeit with altered boundary conditions that necessitate the use of complex-valued crystal momenta to describe the Bloch eigenfunctions of the periodic potential. Using this mapping, we show that the the complex-valued non-Hermitian band structure is the analytical continuation of the real-valued Hermitian band structure for complex momenta, which is well-studied in the context of surface states of finite crystals~\cite{kohn1959analyticProperties,heine1963surface} and which we make extensive use of. 
We also report preliminary evidence of a reentrant delocalization transition in the presence of both periodic potential and disorder, and explore possible connections with non-Hermitian topological pumps. 

This article is structured as follows: In Sec.~\ref{sec:simulationResults}, we report properties of the equilibrium chain conformations observed in molecular dynamics simulations and numerically demonstrate a localization-delocalization transition. In Sec.~\ref{sec:theory}, we derive the diffusion equation governing the probability density of the chains and map it to the Schr\"{o}dinger equation with a non-Hermitian Hamiltonian. We find the eigenstates of the Hamiltonian in Sec.~\ref{sec:groundstate} and in Sec.~\ref{sec:band} show that the delocalization threshold is captured by a branch point in the complex-valued band structure. We also verify the theoretical prediction with the results from simulations. We report a critical exponent associated with the delocalization transition in Sec.~\ref{sec:exponent}, preliminary results on a system with quenched substrate disorder in Sec.~\ref{sec:disorder}, and finally the relation to topologically quantized currents in Thouless pumps in Sec.~\ref{sec:adiabatic}. We discuss the implications of our results and potential future directions in Sec.~\ref{sec:discussion}.

\section{Simulation results}\label{sec:simulationResults}
We first report the results of Langevin dynamics simulations of a discretized version of the system depicted in~\figref{fig:intro}, in which changes in equilibrium conformations at different filling fractions and force angles are readily visualized.
We simulated thermally fluctuating chains of monomers confined to two dimensions using the open-source molecular dynamics software \texttt{HOOMD-Blue}~\cite{Anderson2020} (see \appref{app:sim} for implementation details). The chains are stretched out via a tension force applied to both ends, and experience a grooved substrate potential. Monomers repel each other with a short-ranged potential, and are linked to neighbors along the chain using stiff harmonic springs, to emulate polymers that cannot cross each other and are free to experience shape fluctuations. The substrate potential per unit length, $V(x)$, is periodic along the $x$ direction and constant along the $\tau$ direction (shaded background in \figref{fig:intro}). The tension force, $F$, is applied to the ends of each chain at a specified force angle, $\theta$, with respect to the grooves of the potential (the $\tau$ axis). The force angle is kept constant during each simulation run and quantifies the degree of shear experienced by the polymers. The effect of a finite temperature is incorporated by including viscous drag and introducing random forces on monomers whose strength is related to the desired temperature via a fluctuation-dissipation relation. After an equilibration period, monomer positions can be aggregated over statistically independent time points to obtain equilibrium density profiles of the fluctuating chains. 

\begin{figure}
\centering
\includegraphics{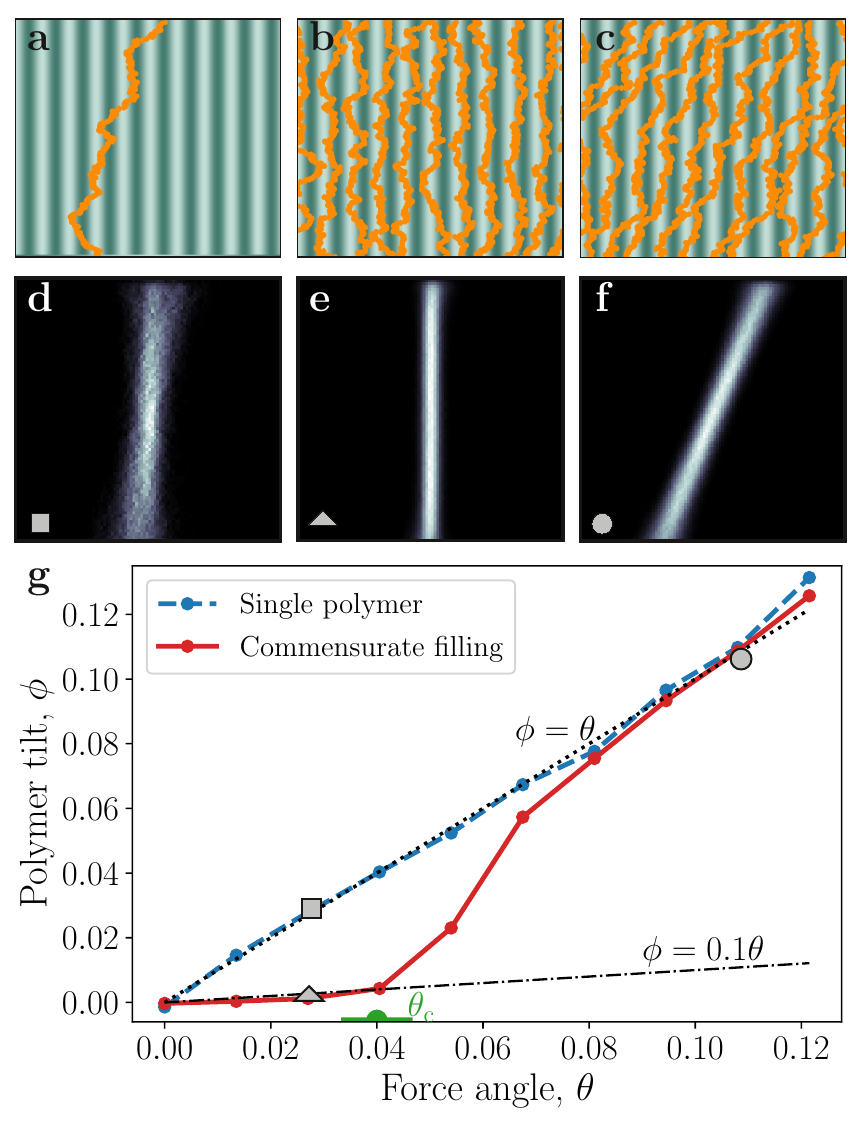}
\caption{\subfiglabel{a--c}, Snapshots of Langevin dynamics simulations of discretized polymer chains (points) on a periodic substrate potential (colormap same as in \figref{fig:intro}). Each chain has equilibrium length $40a$ and the simulation box has width $10a$; $x$ and $\tau$ directions have different scales.
\subfiglabel{a}, Single polymer under low shear. 
\subfiglabel{b, c}, Noncrossing polymers under commensurate filling (one chain per repeating unit of the substrate potential) under low (\subfiglabel{b}) and high (\subfiglabel{c}) shear.
\subfiglabel{d--f}, Aggregated density profiles of equilibrium chain conformations from simulations \subfiglabel{a--c}.
\subfiglabel{g}, Tilt angle of the aggregated polymer conformations, $\phi$, as a function of force angle, $\theta$, as measured in simulations of a single polymer and of multiple polymers under commensurate filling. Gray symbols indicate the parameter values for panels \subfiglabel{a--f}. Dotted line shows $\phi = \theta$. Critical force angle $\tcrit$ is estimated as the intersection of measured tilt--angle curve with $\phi = 0.1\theta$ (dash-dotted line; see \appref{app:measurement} for details). The spacing of simulated $\theta$ values provides the uncertainty in the estimate. From the commensurate curve, we obtain $\tcrit = 0.040 \pm 0.007$ (green symbol on $\theta$ axis).
}
\label{fig:sims}
\end{figure}

When the potential energy experienced by a single monomer is comparable to the thermal energy scale $k_BT$, a single polymer chain wanders across the simulation box with no preferred position [\subfigref{fig:sims}{a}]. Upon subtracting the center-of-mass motion of the chain from the monomer positions at each time step, the equilibrium density profile obtained by aggregating monomer positions over thousands of independent time steps (see \appref{app:sim} for details) displays an overall tilt in the direction of the transverse shear force, as seen in \subfigref{fig:sims}{d}. The tilt angle $\phi$, extracted from the difference in the average $x$ positions from the density profiles near opposite ends of the chain, is seen to align with the force angle $\theta$ (\subfigref{fig:sims}{g}). This alignment shows that a single chain is free to tilt in response to the external tension and is not significantly confined by the substrate potential. 

The polymer conformations are markedly different when the system is prepared at a commensurate filling of one polymer per groove of the periodic potential. At small force angles, the wandering of polymers in the multichain system is suppressed and each chain is localized to a distinct groove [\subfigref{fig:sims}{b}]~\cite{pedro2019protection} breaking ergodicity~\cite{luitz2017ergodic, palmer1982ergodicity}. The aggregated density profile shows that the chains remain vertically aligned even at nonzero shear, except for a small amount of bending near the ends [\subfigref{fig:sims}{e}]. Only at force angles larger than a threshold value, $:=\tcrit$, do the chain conformations acquire a significant tilt along the entire length of the chain [\subfigref{fig:sims}{f}]. The threshold force angle is identified by a sharp increase both in the magnitude of the tilt $\phi$ and in the slope of the $\phi$-$\theta$ relationship (numerical estimation of $\tcrit$ is discussed in \appref{app:measurement}). At force angles much larger than the threshold, the chains align with the applied force [$\phi \approx \theta$, \subfigref{fig:sims}{g}]. At $\theta > \tcrit$, not only do the chains display an abrupt tilt, they also drift back and forth across the substrate at equilibrium with no preferred center-of-mass location. This motion occurs via the diffusion of kinks that carry a chain over a potential peak to the adjacent valley [\subfigref{fig:sims}{c}]; the kink positions and motion must be coordinated across all chains to satisfy the noncrossing constraint~\cite{lehrer1998mott}.

In summary, the commensurate system exhibits two distinct equilibrium phases in simulations: a localized phase with untilted chain conformations confined to individual potential grooves, and a delocalized phase in which chains are tilted in the direction of the applied force and wander across the substrate. 
To explain these phases, we map the density profiles of the classical equilibrium system to the quantum probability evolution of noninteracting fermions experiencing a periodic potential~\cite{pedro2019protection} in the presence of a non-Hermitian term due to shear~\cite{hatano1996localizationprl}. The mapping has its roots in field-theoretic descriptions of the statistics of polymer melts~\cite{Edwards1965,DeGennes1969,Fredrickson2002,matsen2005selfConsistent}, but is simplified by the absence of self-interactions of each chain with itself---the tension is assumed to be large enough that each chain conformation is described by a single-valued function $x(\tau)$ and chains do not double back on themselves along the $\tau$ direction. This feature, which distinguishes \emph{directed} polymer models from regular polymers, allows the corresponding quantum system to be exactly solved for some types of inter-polymer interactions~\cite{gennes1968soluble,rocklin2012constraints}. 

\section{Theoretical analysis}
\subsection{Classical-quantum mapping}\label{sec:theory}

The chains of monomers depicted in \subfigref{fig:sims}{a--c} are rough at a microscopic scale set by the monomer size $\ell_0$. Meanwhile, the potential energy varies at a scale set by the lattice constant $a$, which can be much larger than the monomer size. In the limit that $a \gg \ell_0$, the lattice-scale features of the chain conformations can be described using a coarse-grained model in which monomers are aggregated into mesoscopic segments. Each segment is small enough such that its local environment is more or less homogeneous, but large enough that its fluctuations obey Gaussian statistics~\cite{matsen2005selfConsistent}. This coarse-graining effectively smoothens out the monomer-scale roughness (as depicted schematically in \figref{fig:intro}) burying microscopic degrees of freedom in a multiplicative constant for the partition function (additive constant for the entropy) which does not affect the equilibrium properties.

For each polymer chain, labeled by the index $1 \leq n \leq N$, the instantaneous coarse-grained conformation is then specified by a smooth function $x_n(\tau)$ with $\tau \in [0, L]$ where $L$ is the length of each polymer. The total energy, at small angles $\theta \ll 1$ and small chain slopes $\partial_\tau x_n \ll 1$, is~\cite{rocklin2012constraints, hwa1993disorder, nelson1993boson, lehrer1998mott}  
\begin{equation}\label{energy}
    E = \sum_{n=1}^N \int_0^L d\tau \bigg ( \frac{F}{2} (\partial_{\tau}x_n - \theta)^2 + V(x_n) +  \sum_{n' \neq n}  |c|\delta(x_n - x_{n'}) \bigg).
\end{equation}
The first term  in the integrand captures the energy cost of the chain deviating from a straight line aligned with the force direction. For the coarse-grained description to hold, the transverse fluctuations due to thermal energy at the monomer scale must be small compared to the lattice spacing. The energy associated with a deflection of order $\delta$ over a length $\ell_0$ is order $F(\delta/\ell_0)^2 \times \ell_0 \sim F \delta^2/\ell_0$. The typical deflection $\delta_\text{th}$ due to thermal fluctuations is obtained by balancing this energy against the thermal energy scale $k_BT$, which gives $\delta_\text{th} \sim \sqrt{k_BT \ell_0/F}$. In all our simulations, parameters are chosen such that $\delta_\text{th} \ll a$. 

The second term in \eqnref{energy} implements the position-dependent substrate potential, where $V(x)$ is the potential energy per unit length of the chain experienced at position $x$. The coarse-grained description in the theory and the microscopic description in simulations are matched by setting $V = V_\text{m}/\ell_0$, where $V_\text{m}$ is the potential energy experienced by each monomer.

The last term in \eqnref{energy} incorporates interactions among chains, which are assumed to be entirely local so that chain segments interact only when $x_n(\tau) = x_{n'}(\tau)$~\cite{rocklin2012constraints}. In this work, the only interaction we will consider is that polymers cannot cross each other, which is implemented by taking the limit $|c| \to \infty$. As pointed out by de\,Gennes~\cite{gennes1968soluble}, for noncrossing polymers we can use Girardeau's mapping~\cite{girardeau1960mapping} to eliminate the interaction term from the energy and absorb its effect into the boundary conditions of the probability density functions describing the polymers. We will exploit this feature in the theoretical treatment below (see \eqnref{singlebody}). In simulations, the noncrossing condition is implemented by including extremely stiff contact forces among monomers whose radius is $\ell_0$, so that monomers cannot pass through the gaps between pairs of monomers on other chains. Prior work on a related system~\cite{pedro2019protection} showed that when $\ell_0$ is small, the statistics of such monomer chains quantitatively match theoretical expectations from idealized noncrossing lines. 

We now derive a Schr\"odinger-like equation governing the probability weights associated with the chain conformations at thermal equilibrium. Our approach augments the continuum treatment of shear-free ($\theta=0$) directed polymers in Refs.~\onlinecite{gennes1968soluble,rocklin2012constraints} to include the effects of a periodic substrate potential~\cite{pedro2019protection} and of shear forces following the Hatano-Nelson model~\cite{hatano1996localizationprl,hatano1997vortexPinning}. Consider the conformation of any one of the $N$ polymer chains, $x(\tau')$, and assume it is pinned at two points, $\tau' = 0$ and $\tau' = \tau$, so that $x(0) = x_0$ and $x(\tau) = x_\tau$. The energy of the fragment between the points is denoted by $E[x; 0, \tau]$. The partition function of this fragment can be written as a path integral~\cite{wiegel1986path} over all paths obeying the pinning constraints,
\begin{equation} \label{eq:pathintegral}
    \Psi(x_\tau, x_0, \tau) = \int_{(x_0,0)}^{(x_\tau, \tau)} \mathcal{D}x \exp({-\beta E[x; 0, \tau]}),
\end{equation}
where $(\beta = 1/k_B T)$. By considering the change in the partition function between the vertical coordinates $\tau$ and $\tau + \epsilon$ in the limit $\epsilon \to 0$ (see details in \appref{app:diffusion}), we find that it satisfies the following diffusion equation:
\begin{align}
  \frac {\partial \Psi(x, \tau)}{\partial \tau}  &= \bigg (\frac{1}{2\beta F} \frac{\partial^2}{\partial x^2} - \theta\frac{\partial}{\partial x} - \beta V(x) \bigg ) \Psi(x, \tau), \label{diffusion}
\end{align}
where the linear differential operator in the brackets will take the role of a Hamiltonian upon mapping to quantum mechanics.
The partition function  $\Psi(x_\tau, x_0, \tau)$ with some modifications will help us retrieve the probability distribution $p(x,\tau)$ of finding the point $\tau$ along the chain at horizontal coordinate $x$.

The transformation from \eqnref{eq:pathintegral} to \eqnref{diffusion} is formally similar to the transformation from the Feynman path integral,$\int \mathcal{D}x \exp({i S[x]}/\hbar)$, to the Schr\"odinger formalism in quantum mechanics.  Indeed, if we redefine variables by mapping $\tau = it$, $\beta = \frac{1}{\hbar}$, $F = m$, and $F\theta = g$, \eqnref{diffusion} maps to the time-dependent Schr\"{o}dinger equation,
\begin{align}
  i\hbar \frac {\partial \Psi(x, t)}{\partial t} &= \bigg( \frac{(p + ig )^2}{2m} + \frac{g^2}{2m} + V(x) \bigg ) \Psi(x, t) \label{eq:schrod} \\
  &\equiv \bigg( \Hhn + \frac{g^2}{2m} \bigg) \Psi(x,t),
\end{align}
where $p = -i\hbar \frac{\partial}{\partial x} = -\frac{i}{\beta} \frac{\partial}{\partial x}$.
The nonzero shear component of the forces on the chains manifests itself as an imaginary vector potential $ig=F\theta$~\cite{hatano1996localizationprl}. We identify the Hamiltonian to be $\Hhn$ shifted by a constant, $\frac{g^2}{2m}$; where $\Hhn$ is the continuum version of the periodic Hatano-Nelson Hamiltonian which has been well-studied in the tight-binding limit~\cite{hebert2011hatanoNelson}.

The procedure can be repeated for the full many-body system with the path integral now involving all possible conformations of the $N$ chains. The mapped many-body quantum system then has a Hamiltonian~\cite{hatano1997vortexPinning,rocklin2012constraints},
\begin{equation}\label{hamiltonian}
    \Hmb = \sum_{n=1}^N\bigg( \frac{(p_n + ig )^2}{2m} +\frac{g^2}{2m}  + V(x_n) + \sum_{n' \neq n} |c|\delta(x_n - x_n')\bigg ),
\end{equation}
where $p_n = -i\hbar \frac{\partial}{\partial x_n} = -\frac{i}{\beta} \frac{\partial}{\partial x_n}$. This Hamiltonian describes a one-dimensional system of $N$ quantum particles in a periodic potential $V(x)$ acted upon by a common imaginary vector potential $ig$. Now, we impose the restriction $|c| \to \infty$, which corresponds to the noncrossing interaction. In this limit, the interaction term can be absorbed into the boundary conditions of the many-body wave function~\cite{gennes1968soluble} using Girardeau's mapping~\cite{girardeau1960mapping} which effectively maps bosons with contact repulsion to \emph{non}-interacting fermions. By considering only fermionic many-body wavefunctions, the noncrossing condition is automatically satisfied. The many-body Hamiltonian then becomes a sum of single-body terms,  
\begin{equation}\label{singlebody}
    \Hmb = \sum_{n=1}^N\bigg( \mathcal{H}_n(g) +\frac{g^2}{2m}\bigg ).
\end{equation}
The statistics of $N$ noncrossing, fluctuating lines has been mapped to a quantum mechanical problem of noninteracting fermions, each experiencing the same periodic scalar potential $V$ and constant  imaginary vector potential $ig$. 

\subsection{Eigenstates and ground-state dominance}\label{sec:groundstate}

Besides simplifying the description of many-body densities which automatically satisfy the noncrossing condition, the mapping to quantum mechanics motivates the use of a spectral expansion to represent the solution to \eqnref{diffusion}. A general solution to a linear differential equation such as \eqnref{diffusion} can be written as a superposition of eigenfunctions $\Psi_m(x)$ of the Hamiltonian, 
\begin{equation}
    \Psi(x, \tau) = \sum_m c_m e^{-\beta (\varepsilon_m + \frac{F\theta^2}{2}) \tau} \Psi_m(x),
\end{equation}
where
\begin{equation}
  \label{eigenequation}
  \Hhn \Psi_m(x) = \bigg( \frac{(p + ig)^2}{2m} + V(x) \bigg )\Psi_m(x)= \varepsilon_m \Psi_m(x)
\end{equation}
for some quasienergy eigenvalue $\varepsilon_m$, and $c_m$ are constants fixed
by the initial condition $\Psi(x,0)$.
We will index the eigenfunctions in increasing order of the real part of the quasienergy, $\rea \varepsilon_{m^\prime} > \rea \varepsilon_{m}$ for $m^\prime > m$. As the coordinate along the polymer increases from $\tau = 0$, the amplitudes of eigenstates relative to the ``ground state'' ($m=0$) decay exponentially as $e^{-\beta \rea(\varepsilon_{m} - \varepsilon_0)\tau}$. Far from the boundary, the polymer's profile is dominated by the ground-state wave function of the time-independent Hamiltonian with the lowest real component of $\varepsilon_m$, $$\Psi(x, \tau) \sim e^{-\beta (\varepsilon_0+ \frac{F\theta^2}{2}) \tau} \Psi_0(x),$$ a situation termed \emph{ground state dominance}~\cite{deGennes1979}. Ground state dominance holds in the interior of polymers with lengths that satisfy $L \gg 1/[\beta(\rea \varepsilon_1 -  \rea \varepsilon_0)]$; the lower the real energy gap between the lowest two  quasienergies, the longer the polymer needs to be.

To describe the density profiles of polymers away from the ends, we also need the contribution to the partition function of a polymer being pinned at $x(L) = x_L$ and propagating \emph{downwards}. We will write this partition function contribution as,
\begin{equation}
    \Psit(x_\tau, x_L, \tau) = \int_{(x_L,L)}^{(x_\tau, \tau)} \mathcal{D}x \exp({-\beta E[x; \tau, L]}).
\end{equation}
The diffusion equation obeyed by $\Psit$ is obtained by rotating the coordinate system $(x,\tau) \to (-x,-\tau)$ in \eqnref{diffusion}, leading to
\begin{align}
  \frac {\partial \Psit(x, \tau)}{\partial \tau}  &= \bigg (-\frac{1}{2\beta F} \frac{\partial^2}{\partial x^2} - \theta\frac{\partial}{\partial x} + \beta V(x) \bigg ) \Psit(x, \tau).
\end{align}
On repeating the quantum mapping, and assuming an even potential, $V(-x) = V(x)$, we get,
\begin{align}
  -i\hbar \frac {\partial \Psit(x, t)}{\partial t} &= \bigg( \frac{(p - ig )^2}{2m} + \frac{g^2}{2m} + V(x) \bigg ) \Psit(x, t) \label{eq:schrod-rev} \\
  &= \bigg( \Hhn^\dagger + \frac{g^2}{2m} \bigg) \Psit(x,t).
\end{align}
Note the reversed sign of time. We now expand $\Psit$ using the eigenfunctions of $\Hhn^\dagger$ as $\Psit(x, \tau) = \sum_m \tilde c_m e^{-\beta (\tilde\varepsilon_m+ \frac{F\theta^2}{2}) (L - \tau)} \Psit_m(x)$, where $$\Hhn^\dagger \Psit_m(x) = \bigg( \frac{(p - ig)^2}{2m} + V(x) \bigg )\Psit_m(x)= \tilde\varepsilon_m \Psit_m(x)$$ and the coefficients $\Tilde c_m$ are fixed by the boundary condition at $\tau = L$.  

Although $\Hhn \neq \Hhn^\dagger$ because of the non-Hermiticity induced by a finite force angle, the eigenfunctions and eigenvalues of the Hamiltonians in \eqnref{eq:schrod} and \eqnref{eq:schrod-rev} are closely related.
A diagonalizable non-Hermitian Hamiltonian can be written as~\cite{ashida2020nonhermitian}
\begin{equation}
    H = \sum_n \lambda_n \ket{R_n}\bra{L_n},
\end{equation}
where the right eigenstates, $\ket{R_n}$, and the left eigenstates, $\bra{L_n}$, form a biorthonormal basis, $\expect{L_i|R_j} = \delta_{ij}$. By taking the conjugate transpose of the above equation, we get $$H^\dagger = \sum_n \lambda^*_n \ket{L_n}\bra{R_n}.$$ We identify $\Psi_m(x) = \expect{x|R_m}$ and $\Psit_m(x) = \expect{x|L_m}$ such that $\tilde \varepsilon_m = \varepsilon_m^*$ and $\int \! dx \, \Psit^*_m(x)\Psi_n(x) = \delta_{mn}$. Now, since $\Hhn$ (as well as $\Hhn^\dagger$) is real-valued, its eigenstates are either real with real eigenvalues, or come in complex-conjugate pairs. Using this, we choose to index the eigenstates $\Psit_m(x)$ such that,
\begin{equation}
    \int_0^{L_x} \! dx \, \Psit_m(x)\Psi_n(x) = \delta_{mn}\quad \text{and}\quad \tilde \varepsilon_m = \varepsilon_m
\end{equation}
In particular, the ground state has a real eigenvalue even for the non-Hermitian problem~\cite{hatano1997vortexPinning}, and far from the upper end of the polymer we obtain
$$\Psit(x, \tau) \sim e^{-\beta (\varepsilon_0+ \frac{F\theta^2}{2}) (L-\tau)} \Psit_0(x).$$

The polymer density is expressed in terms of the two partition function contributions as
\begin{equation}
  \label{eq:density}
  p(x,\tau) = \frac{1}{Z}\Psi(x,x_0,\tau)\Psit(x,x_L,\tau),
\end{equation}
where $Z = \int \! dx  \, \Psi(x,x_0, \tau)\Psit(x,x_L, \tau) = \Psi(x_L,x_0,L)$ is the full partition function of the chain with end points $(x_0,0)$ and $(x_L,L)$ and is therefore independent of the $x$ and $\tau$ coordinates. While $p(x,\tau)$ can be expanded in terms of the eigenfunctions $\Psi_m(x)$ and $\Psit_m(x)$, we operate in the limit of long polymer chains where the density far from the ends is dominated by the ground state:
\begin{equation}
  \label{eq:density-groundstate}
  p(x,\tau) \sim \Psi_0(x)\Psit_0(x).
\end{equation}

These quantities are readily translated to the corresponding many-body quantities. Using Girardeau's mapping~\cite{girardeau1960mapping}, the many-body eigenstate, $\Psi(\bs x) = \Psi(x_1, x_2, ..., x_N)$, of the Hamiltonian in \eqnref{hamiltonian} is the Slater determinant of the single-body wave functions~\eqnref{singleBodyWF} (see \appref{sec:manybody} for details). The Slater determinant ensures that $\Psi(x_1, x_2, ..., x_N) = 0$ whenever any $x_i = x_j$, thus enforcing the noncrossing condition. The associated energy, is the sum of the single-particle eigenenergies of states in the Slater determinant, $\bs{\varepsilon} = \sum_i \varepsilon_i$. The many-body ground state, which determines the polymer profiles away from the ends, is therefore the Slater determinant of the lowest $N$ single-particle eigenstates. The analogue of \eqnref{eq:density-groundstate} for  the many-body probability density over the polymer chain coordinates $\bs x = (x_1, x_2, \dots, x_N)$ at any $\tau$ as,
\begin{equation}\label{manybodyprob}
    p(\bs x;\tau) = \frac{1}{Z}\Psi(\bs x, \tau)\Psit(\bs x, \tau) \sim \Psi_0(\bs x)\Psit_0(\bs x).
\end{equation}

\subsection{Imaginary gauge transformation and the delocalization transition}

The problem of finding the polymers' density profile has reduced to finding the lowest $N$ eigenstates of the Hamiltonian~\eqnref{eigenequation} (ordered by the real part of the quasienergy). To do so, we use the fact that if $\Psi'(x)$ is an eigenstate of the shear-less Hamiltonian, $\mathcal H(g=0) = \frac{p^2}{2m} + V(x)$ with eigenvalue $\varepsilon$, then $\Psi(x) = e^{\frac{gx}{\hbar}}\Psi'(x) =e^{F\beta \theta x}\Psi'(x)$ is an eigenstate of $\Hhn$ with the same eigenvalue. (This is analogous to a gauge-transformation of the vector-potential~\cite{aharanov1959bohm}.)

When $V(x)$ is periodic, Bloch's theorem applies and the eigenfunctions $\Psi'_k(x)$ should be of the form $e^{ikx}u_k(x)$ with $u_k(x)$ having the same periodicity as the potential, i.e., $u_k(x+a) = u_k(x)$. However, to ensure that $\Psi(x)$ is physical (in particular that it obeys periodic boundary conditions) we must choose $k$ to be complex such that $\ima(k) = g/\hbar = F\beta\theta$ to cancel out the `gain factor' $e^{\frac{g}{\hbar}x}=e^{F\beta \theta x}$. The single-particle, normalizable eigenstates of $\Hhn$ are then
\begin{equation}\label{singleBodyWF}
    \Psi(x) = \Psi_k(x) = e^{i\rea(k)x}u_k(x),
\end{equation}
with $k = \rea(k) + iF\beta \theta$. Such Bloch waves with complex $k$ have been used to describe the evanescent surface states of a finite crystal~\cite{heine1963surface, pendry1970bound, inglesfield1982surface} and more recently to elucidate the non-Hermitian skin effect~\cite{yokomizo2019nonbloch, yao2018edge}.

The essential physics of the nonzero tilt angle for polymer conformations is captured in the localization, or lack thereof, of the fermionic ground states constructed from the Bloch waves. A superposition of $N$ Bloch waves is generically delocalized through the whole lattice and has equal weight on all unit cells. The equilibrium density profile of the polymer far from the ends is therefore uniform across the system: the polymer wanders freely and visits each groove with equal probability over long times. The wandering polymer aligns its conformation to the force angle to minimize its free energy, leading to a conformational tilt which grows with tilt angle as observed in simulations.

An exception to the generic delocalized state occurs under commensurate filling of one polymer per groove of the periodic potential. At zero shear, appropriate superpositions of the Bloch waves can be used to construct an alternate basis for the lowest band, consisting of a set of Wannier functions
$$\Phi'_j(x) = \Phi'(x-ja),$$
each centered on the $j$th unit cell of the periodic potential and exponentially localized in the $x$ direction, $\Phi'(x) \sim \exp(-\lambda |x|)$~\cite{kohn1959analyticProperties, brouder2007wannier}. Here, $\lambda>0$ is an inverse localization length determined by features of the \emph{complex} band structure $\varepsilon(k)$.

Once we have identified an exponentially localized set of basis states, the mechanism of how shear causes delocalization can be framed in a very general manner using the imaginary gauge transformation~\cite{hatano1997vortexPinning,hatano1998delocalization}.
Consider a many-body Hermitian (shear-free) system, $\bs H(g=0) = \sum_n \frac{p_n^2}{2m} +V(\bs x)$. The potential energy, $V(\bs x)$, might be either periodic (as in the current problem), or disordered, or both. It may even have possible interaction terms.
Regardless of the localization mechanism, if the many-body ground-state of the shear-free system $\bs \psi'(\bs x; g=0)$ is exponentially localized, the effective eigenstate of each polymer chain dies off away from its mean/typical position as $\Psi'(x) \sim \exp(-\lambda |x|)$ where $\lambda>0$ is the many-body inverse localization length~\cite{Kohn1964insulating, brouder2007wannier}. 
Using the gauge transformation, we see that in the presence of shear, the many-body (right) ground-state is $\bs \psi(\bs x; g) = e^{F \beta \theta \sum_n x_n} \bs \psi'(\bs x; g=0)$. The gauge transform is permissible as long as $e^{F \beta \theta x} \Psi(x)$ is well-behaved, i.e., it still must satisfy the periodic boundary conditions. This condition is met as long as $\lambda>F \beta \theta$. The critical shear at which delocalization happens is then $F\theta_c = \frac{\lambda}{\beta}$ (compare with the conductivity expression in Ref.~\onlinecite{kawabata2021scaling}). If the density profiles of chains in the shear-free  many-body ground state $\bs \psi'(\bs x; g=0)$ fall off faster than exponentially with distance from the mean chain position [for example, as a Gaussian profile $\psi^\prime(x) \sim \exp(-\lambda^2 x^2)$ which decays faster than any exponentially decaying function at large enough $x$], then no amount of shear will delocalize the system. By contrast, if the localization is weaker than an exponential falloff (for example, as a power-law decay with distance from the mean position), then an infinitesimal amount of shear is sufficient to delocalize the system.

Finding the localization length in the Hermitian problem is, thus, equivalent to finding the critical shear in the non-Hermitian problem. Conversely, finding the critical shear, either experimentally or theoretically, in the non-Hermitian system is equivalent to finding the localization length in an exponentially localized Hermitian system. In our specific system, this localization length is precisely that of the Wannier functions at commensurate filling, which we explicitly calculate in Sec.~\ref{sec:band}. Interestingly, the effect of the shear force on the polymer density profiles themselves is minimal in the localized phase. Observing that the right eigenstate in the non-Hermitian system falls off as $\sim e^{F \beta \theta x} \exp(-\lambda |x|)$, the inverse localization length in the sheared system appears to be $\lambda - F\beta \theta$. This is, however, the inverse localization length of the right eigenstate only. The polymer's probability profile consists of the product of both the right and the left eigenstate [\eqnref{eq:density}]. The left eigenstate of $\Hhn$ is the right eigenstate of $\Hhn^\dagger = \mathcal{H}(-g)$ which suffers an opposite gauge transformation. The inverse localization length of the polymer density profiles under shear, obtained by multiplying the left and right ground states [\eqnref{eq:density-groundstate}], is then the same as that of a system without shear ($\theta=0$). Away from the polymer ends, non-Hermiticity does not  affect polymer density profiles in the localized phase~\cite{hatano1998delocalization}.

We note that the gauge transformation of a localized non-Hermitian system to a localized Hermitian system is not possible in a tight-binding (discrete) model such as the lattice-based Hatano-Nelson model with a periodic potential~\cite{hebert2011hatanoNelson}. Indeed, there is no similarity transformation that maps a generic non-Hermitian matrix with complex eigenvalues to a Hermitian matrix. A key difference between discrete and continuum models is that boundary conditions are part of the operator itself in the former but not in the latter. For the continuum model, we have to adjust the boundary conditions of the mapped Hermitian system to ensure the original non-Hermitian system has periodic boundary conditions. When the many-body non-Hermitian phase is localized, the Hermitian phase is localized as well, and we can retain periodic boundary conditions ensuring the mapped Hamiltonian is Hermitian as well as self-adjoint. In the delocalized phase, we are forced to place special nonperiodic boundary conditions in the mapped Hermitian problem such that the operator is no longer self-adjoint. A continuum Hermitian operator with an infinite-dimensional Hilbert space can have complex eigenvalues if it is not self-adjoint~\cite{ahari2016selfadjointness,capri1977selfadjointness}; this fact explains the complex eigenenergies in our system even though it can be mapped to a Hermitian Hamiltonian. The gauge transformation also allows us to make use of the well-studied properties of the analytical continuation of the Hermitian spectrum~\cite{kohn1959analyticProperties} which is essentially the spectrum of the Hermitian operator under modified (nonperiodic) boundary conditions.

Having identified the mechanism for delocalization in the system under commensurate filling, we now turn to analyzing the energy spectrum, $\varepsilon(k)$, to determine the value of the critical force angle; we also compute the tilt angle of the polymers from the corresponding eigenfunctions.

\begin{figure}[tb]
	\includegraphics{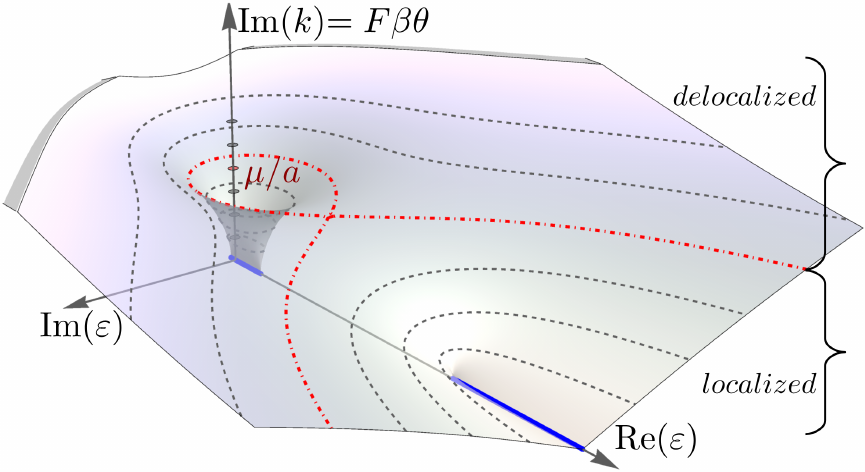}
	\caption[]{\label{Fig:bandStructure} 
	The (complex-valued) energies, $\varepsilon$, of the Hamiltonian $\frac{p^2}{2m} + V(x)$ , where $V(x)$ is periodic, as a function of the imaginary component of the Bloch wave-vector, Im$(k)$, the value of which is set by the shear force $F\theta$ and temperature $\beta$ in the polymer system. When Im$(k) = 0$, the energies are real and form separate bands (shown in blue). As the shear force is increased Im$(k)$ increases and the energies form complex-valued ovals (grey dotted contours). At the critical value of Im$(k) = \mu/a$ the ground-state oval meets the first excited band and a commensurate filled crystal is no longer a band insulator. This is the delocalization mechanism exhibited by the polymer system.  While the complex energies shown here have been computed for the specific potential $V(x) = V_0\cos(2\pi x/a)$, with $V_0 = 1$ (see \appref{app:computation} for computation details), this behaviour is generic for even one-dimensional potentials with $V(x) = V(-x)$ and are expected to hold with some modifications for nonsymmetric periodic potentials as well~\cite{kohn1959analyticProperties,prodan2006analytical}.}
\end{figure}

\subsection{Non-Hermitian gap closure and critical shear}\label{sec:band}

At finite shear, the quasienergies of the eigenstates become complex-valued since the Hamiltonian is non-Hermitian, $\Hhn^\dagger = \mathcal{ H}(-g) \neq \Hhn$. The complex energy bands, $\varepsilon_n(k)$, can be regarded as distinct Riemann sheets over the complex $k$ plane of a multivalued complex function $\varepsilon(k)$~\cite{kohn1959analyticProperties, heine1963surface,prodan2006analytical} with each sheet corresponding to a particular band. (When used with an argument $k$, the expression $\varepsilon_n(k)$ denotes the $n$th energy band associated with the periodic potential; bands are ordered according to the real part of the energy.) While the bands are separated for real $k$, adjacent sheets meet at branch points of  $\varepsilon(k)$ which occur at complex $k$: The $n$th sheet meets the $(n+1)$th sheet at wave-vector values $k_n = \pm \frac{\pi}{a} \pm i \frac{\mu_n}{a}$, where the dimensionless numbers $\mu_n$, which quantify the distances of the branch points from the real $k$ axis, are determined by the potential function $V(x)$. The values of $\mu_n$ will determine the value of the critical shear in our system.

In the polymer model, commensurate filling ensures that the quantum particles completely fill the groundstate energy band $\varepsilon_0(k)$. At $F\theta = 0$, when there is no shear, the energies are real and there is a finite gap between the groundstate band and the higher band. At nonzero shear force $F\theta$, the bands become complex-valued and turn into ovals in the complex plane. As $F\theta$, and therefore $\ima(k)$, increases, these oval energy bands grow in size (\figref{Fig:bandStructure}) and the separation between bands $\varepsilon_0(k)$ and $\varepsilon_1(k)$ is reduced as the branch point at $k_0$ is approached (\figref{Fig:bandStructure}). When the shear force equals the branch point value, $F\beta\theta = \mu_0/a$, the ground state energy band meets the higher band. The energy gap closes and the system at commensurate filling is now a conductor that enables probability flows driven by the imaginary vector potential. These probability flows manifest themselves as tilts in the polymer density profiles, \figref{fig:sims}.

\begin{figure}[tb]
	\includegraphics{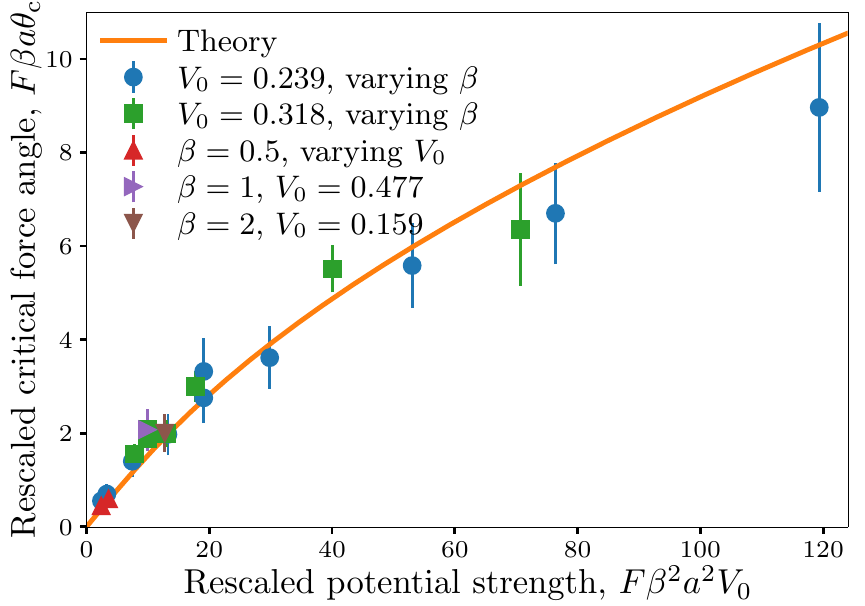}
	\caption[]{\label{Fig:phase}
Critical force angles measured from simulations with different potential amplitudes and temperatures, rescaled by the energy and force scales governing the underlying Schr\"odinger equation
[\eqnref{dimensionlesseqn}]. Symbols are labeled according to the parameter which is kept constant in distinct sets of simulations. The other parameters are $F=20$, $a=1$,
$N=10$ in simulation units. Error bars show uncertainty in the critical force estimate due to the finite sampling resolution of applied shear forces. Solid curve shows the theoretical prediction,
     \eqnref{eq:critical_theory}.
	}
\end{figure}

The critical angle of force, $\tcrit$, at which the polymers acquire a collective tilt (\figref{fig:sims}) is therefore predicted to be
\begin{equation}
  \label{eq:critical_theory}
  \tcrit = \frac{\mu_0}{F\beta a}. 
\end{equation}
Notably, the value of $\mu_n$ is sensitive to the details of the continuous periodic potential; its dependence on the energy gap or the amplitude of the potential energy is nonuniversal~\cite{goedecker1999scaling, ismail1999locality}. This subtlety is not captured by tight-binding studies of the complex band structure~\cite{hebert2011hatanoNelson}, which gloss over the details of the periodic potential and suggest that $\tcrit$ should scale with the energy gap in a universal manner.

One can transform $x$ to the dimensionless coordinate $r = x/a$ (where $a$ is the lattice constant) in \eqnref{eigenequation}. Upon rescaling the wavefunction to a new variable $\Psi_n'(r) \equiv \Psi_n(r) e^{-F\beta\theta a r}$ we get,
\begin{equation}\label{dimensionlesseqn}
    -\frac{1}{2 }\frac{\partial^2}{\partial r^2} \Psi'_n(r) + V_0F\beta^2 a^2 V'(r) \Psi'_n(r) = \varepsilon_n F\beta^2 a^2 \Psi'_n(r),
\end{equation}
where $V_0$ is the amplitude of the periodic potential and $V'(r) \equiv V(r)/V_0$. For a particular functional form of the rescaled potential $V'(r)$, the system is then governed by two dimensionless quantities: the dimensionless shear force, $F\beta \theta a$ and the generalized potential strength, $V_0F\beta^2 a^2$. In \figref{Fig:phase} we report measurements of $\tcrit$ from molecular dynamics simulations in which the parameters $\beta$ and $V_0$ were varied for our choice of potential $V'(r) = \cos (2\pi r)$. We find that delocalization thresholds from simulations covering a broad range of parameter values collapse onto a narrow region in the force angle-potential strength plane when rescaled according to the quantum mapping. Furthermore, the rescaled critical force angles are consistent with the theoretical prediction of the localization-delocalization transition---the branch point distance $\mu_0$ at the given potential amplitude [\eqnref{eq:critical_theory}]. The agreement does not involve any fitting parameters, and is robust to changes in the numerical estimation of the critical force angle from simulations (see \appref{app:measurement}). 

Although we only considered a system with commensurate filling in our simulations, the non-Hermitian band structure also determines the delocalization behavior for other filling densities $f = N/M$ (where $M = L_x/a$ is number of unit cells) of the polymers. When $f$ is
noninteger the polymers are expected to be generically delocalized because there is no energy gap separating the last occupied single-particle state from the first unoccupied state. By contrast, when $f$ is an integer the system exhibits a transition from a localized to delocalized state at a critical force angle given by $\frac{\mu_f}{F\beta a}$.

The delocalization transition due to closure of the non-Hermitian energy gap is also apparent in the behaviour of the wave functions for filled bands. We can write the many-body wave function of the commensurate system as a real-valued Slater determinant of Wannier functions defined on the ground state band (see \appref{sec:manybody}). The Wannier functions of the $n$th band, at zero shear, are also known to depend on the branch point locations $\mu_n$: their spatial profiles fall off as $\sim\exp(-\mu_n x/a)$ at large $x$~\cite{kohn1959analyticProperties, brouder2007wannier}. 
Delocalization of the wavefunctions is then linked to the threshold at which the ``gain factor" $e^{\frac{gx}{\hbar}}$ exceeds the falloff $e^{\frac{\mu_n x}a}$, recovering the prediction for the critical angle, $\mu_n/a = g_c/\hbar = F\beta\tcrit$.

The exponential falloff of the Wannier functions is known to include a power-law prefactor~\cite{he2001wannier}: the complete functional form for the large-$x$ behavior is $\Phi'(x)\sim x^{-\alpha}\exp(-\mu_n x/a)$ for the shear-free system. Right at the transition point, the gain factor cancels the exponential decay and it is the power-law falloff that survives. The exponent for the falloff, $\alpha$, is derived from the analytic properties of $\varepsilon(k)$ for 1D periodic potentials,  is known to be universal, and equals $3/4$~\cite{he2001wannier}. This feature provides a critical exponent for the phase transition---at the critical point, the polymers' probability density in the bulk is expected to fall off as $\rho(x)\sim [e^{\frac{g_c x}{\hbar}}\Phi'(x)]^2\sim x^{-3/2}$, independently of the details of the periodic potential. 

\subsection{Critical exponent of diverging shear modulus}\label{sec:exponent}
In the vicinity of the branch points of the energy surface, the imaginary part of the energy is known to vary as as $\varepsilon(k) \sim |\ima(k) - \ima(k_n)|^{1/2}$~\cite{heine1963surface}. This dependence suggests a universal divergence in the shear response of the system near the critical force angle. We can define the shear modulus, $\Gamma$, of the system as the applied stress, $F\theta/a$, divided by the resulting strain, $\phi$. \figref{fig:sims} shows that for a commensurate system, $\Gamma=F\theta/(a\phi)$ equals $F/a$ at $\theta\gg \tcrit$. As $\theta$ approaches $\tcrit$ from above, $\phi$ decreases faster than $\theta$ so that $\theta/\phi$, and hence $\Gamma$, diverges as $\theta \to \tcrit$. For commensurate filling we therefore expect, 
\begin{equation}
    \phi \propto \begin{cases}
    0 &\mbox{if } \theta \ll \tcrit, \\
    (\theta - \tcrit)^\eta & \mbox{if } \theta \sim \tcrit,\\
    \theta &\mbox{if } \theta \gg \tcrit, \end{cases}
\end{equation}
for some critical exponent $\eta$ which we find to be $\frac{1}{2}$ in the following.

The tilt of the polymer chains, $\phi$, can be measured by the averaged slope of the polymer configurations,
\begin{equation}\label{tilter}
    \phi(\tau) \sim \overline{\langle \partial_\tau  x_n \rangle}=  \frac{1}{N}\sum_{n=1}^N\int_0^{L_x} \! d\bs{x} \, x_n \frac{\partial p(\bs{x},\tau)}{\partial\tau},
\end{equation}
where the overline denotes averaging over the $N$ polymers. The tilt, $\phi(\tau)$, depends on the coordinate $\tau$ because of the influence from the initial conditions at $\tau=0$ and $\tau=L$. Indeed in \subfigref{fig:sims}{e,f} we see that the polymers' density profile shows a small amount of bending near the ends. We will be interested in the value of $\phi$ in the bulk of the polymer system, that is at large values of $\tau$ and $L-\tau$. In this regime, the ground state dominance applies such that $(\Psi,\Psit) \to (\Psi_0,\Psit_0)$ and we find that the local tilt angle, $\phi(\tau)$ becomes a constant (see \appref{app:tilt} and Refs.~\onlinecite{hatano1996localizationprl,hatano1997vortexPinning}),
\begin{equation}\label{eq:tilt-manybody-groundstate}
\phi \sim \overline{\langle \partial_\tau  x_n \rangle} = -\frac{1}{NF} \frac{\partial \bs{\varepsilon}_0}{\partial \theta}.
\end{equation}
Here, $\bs{\varepsilon}_0$ is the many-body groundstate energy which is the sum of the lowest $N$ single-body energies,
\begin{equation} \label{eq:manybodygs}
  \emb_0 = \sum_{i=1}^{N} \varepsilon_i.
\end{equation}

At commensurate filling and large $N$, the sum of single-particle energies that yields the many-body groundstate energy can be written in the form of an integral over the complex-valued energy band $\varepsilon_0(k)$:
\begin{equation}
  \emb_0 = \int_C \! \frac{dk}{2\pi/(Na)} \, \varepsilon_0(k)= \int_C \! \frac{dk}{2\pi/(Na)} \, \rea(\varepsilon_0(k)),
\end{equation}
where the complex contour $C$ is the line segment connecting $k = -\pi/a + i \beta F\theta$ to $k = \pi/a+ i \beta F \theta$ at constant $\ima(k) = \beta F\theta$. In the last step, we used the fact that since the single-body eigenvalues are either real or occur in complex conjugate pairs, the many-body ground state energy for a filled band is guaranteed to be real. 

\begin{figure}[t]
  \centering
  \includegraphics{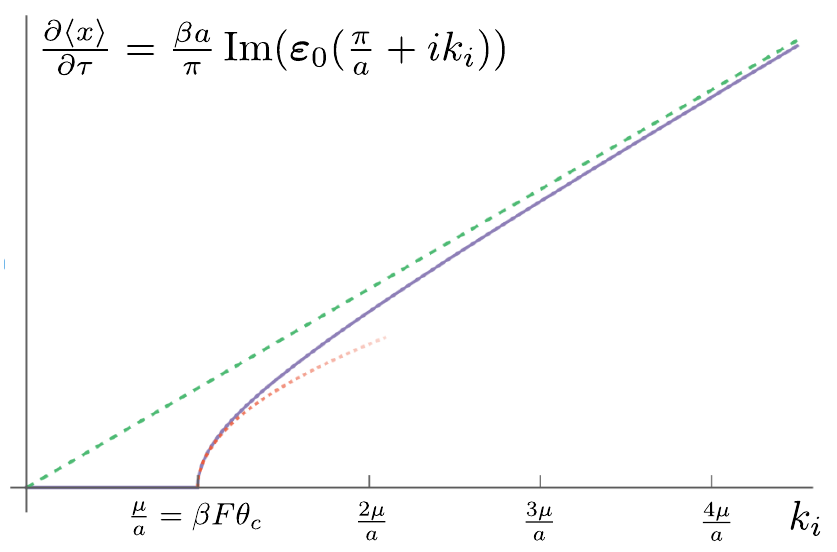}
  \caption{The theoretically predicted $\phi$ vs. $\theta$ graph for a commensurate polymer system. By \eqnref{eq:phi-summary}, the tilt of the polymer chains, at shear value of $F\theta$ and temperature $1/\beta = k_B T$, is given by $\frac{\beta a}{\pi}\ima \varepsilon(k)$ with $k=\frac{\pi}{a} + i F\beta \theta := \frac{\pi}{a} + i k_i$. Shown is the numerically computed value of this expression (blue solid line) for a cosine potential, $V(x) = \cos(2\pi x/a)$. The green dashed line is $\phi = \theta = \frac{k_i}{F\beta}$, to which the tilt asymptotically tends to at large shear; the red dotted curve is $c(k_i-\frac{\mu}{a})^\frac{1}{2}$, demonstrating the critical exponent is $\frac 1 2$ (the value of $c$ is set by fitting). As predicted, the tilt is exactly zero below the critical shear [\eqnref{eq:phi-summary}]. }
  \label{fig:phitheta}
\end{figure}

Using \eqnref{eq:tilt-manybody-groundstate}, the average tilt angle far away from the polymer ends is given by
\begin{align} 
  \phi = -\frac{1}{NF} \frac{\partial{\emb}_0}{\partial{\theta}} &= -\frac{1}{NF}\int_C \! \frac{dk}{2\pi/(Na)} \, \frac{\partial \rea(\varepsilon_0(k))}{\partial \theta} \nonumber \\
  &= -\frac{\beta a}{2\pi} \int_C \! dk \, \frac{\partial \rea(\varepsilon_0(k))}{\partial \ima(k)}. \label{eq:tilt-complex}
\end{align}
Note that the line element $dk$ involves variations only in $\rea(k)$, so we can safely apply the derivative to the integrand. We now use the fact that the complex energy function corresponding to the lowest band, $\varepsilon_0(k)$, is a Riemann sheet of a multivalued function $\varepsilon(k)$ and is analytic everywhere away from the branch points at $k_0=\pm \frac{\pi}{a} \pm i \frac{\mu_0}{a}$ which are encountered only when $\theta = \tcrit$~\cite{kohn1959analyticProperties}. As a result, away from the critical angle the Cauchy-Riemann equations for the analytic function $\varepsilon_0(k)$ can be used to rewrite \eqnref{eq:tilt-complex} as
\begin{align}
  \phi &= \frac{\beta a}{2\pi} \int_C \! dk \, \frac{\partial \ima(\varepsilon_0(k))}{\partial \rea(k)} \nonumber \\
  &= \frac{\beta a}{2\pi}\left[\ima\left(\varepsilon_0\left(\frac{\pi}{a} + i \beta F \theta\right)\right) - \ima\left(\varepsilon_0\left(-\frac{\pi}{a} + i \beta F \theta\right)\right)\right],\nonumber\\
  &=\frac{\beta a}{\pi}\ima\left(\varepsilon_0\left(\frac{\pi}{a} + i \beta F \theta\right)\right). \label{eq:tilt-final}
\end{align}
In the last line we used the fact that $\varepsilon_0(-k^*) = \varepsilon_0^*(k)$ since the Hamiltonian is real~\cite{kohn1959analyticProperties}.

Now for $\theta < \theta_c$, the integration contour $C$ traces a complete loop of the closed oval corresponding to the lowest band in \figref{Fig:bandStructure}, such that $\varepsilon_0(k)$ evaluates to the same value at the endpoints of the contour and the value of the integral, \eqnref{eq:tilt-complex}, is identically zero. For $\theta > \theta_c$, the closed oval corresponding to the lowest band `opens up' by merging with the higher band. The size of the opening along the $\ima(\varepsilon)$ direction is proportional to the acquired tilt via \eqnref{eq:tilt-final}. Near the branch point, $k_0 = \pi/a + i F \beta \theta_c$, it is known that $\ima\left(\varepsilon_0\left(\frac{\pi}{a} + i \beta F \theta\right)\right)$ behaves like $\sqrt{\theta -\theta_c}$ since the branch point is always of order one~\cite{kohn1959analyticProperties}. Finally, for very large values of $\theta$, the momentum term of the Hamiltonian dominates the potential term and we can show that $\ima\left(\varepsilon_0\left(\frac{\pi}{a} + i \beta F \theta\right)\right) = \frac{\pi\theta}{\beta a}$. 

These calculations rely only on the analytic properties of $\varepsilon(k)$ for one-dimensional periodic potentials and on the existence of a gap between the lowest band and higher bands. We stress that while the location of the branch point in the complex momentum plane (i.e., the value of $\mu_0$) depends on system details, its existence and the order of the branch point are independent of the specifics of the periodic potential~\cite{kohn1959analyticProperties,heine1963surface,prodan2006analytical}. Therefore, the critical exponent of $1/2$ is universal for all substrate potentials that are periodic along the $x$ direction and constant along the $\tau$ direction. The details of the calculation linking the branch point structure to the polymer conformations can be found in \appref{app:tilt-analytical} along with figures of the complex band-structure for a cosine potential and a calculation of the tilt for arbitrary filling (see \eqnref{eq:largetheta} in \appref{app:tilt-analytical}).

In summary, we have established that for a commensurate system, the polymer tilt angle is determined by the complex-valued lowest energy band and behaves as
\begin{equation}\label{eq:phi-summary}
    \phi = \frac{\beta a}{\pi}\ima\left(\varepsilon_0\left(\frac{\pi}{a} + i \beta F \theta\right)\right) = \begin{cases}
    0 &\mbox{if } \theta < \tcrit, \\
    c (\theta - \tcrit)^{\frac{1}{2}} & \mbox{if } \theta \gtrsim \tcrit,\\
    \theta &\mbox{if } \theta \gg \tcrit, \end{cases}
\end{equation}
for some constant of proportionality, $c$, which depends on the specifics of the periodic potential. \figref{fig:phitheta} shows this theoretical prediction, numerically computed from the complex band structure of a cosine potential as used in our simulations (solid line). The tilt exhibits the expected functional forms near the critical point (dotted line) and at large force angles (dashed line). 

In the simulations performed for this paper, the polymer chains had a length of $40 a$ while the horizontal length of the simulation box was $10 a$, too limited to extract a critical exponent from the tilt angle measurements. We leave a numerical verification of the predicted divergence exponent for future work.

\begin{figure*}[t]
  \centering
  \includegraphics{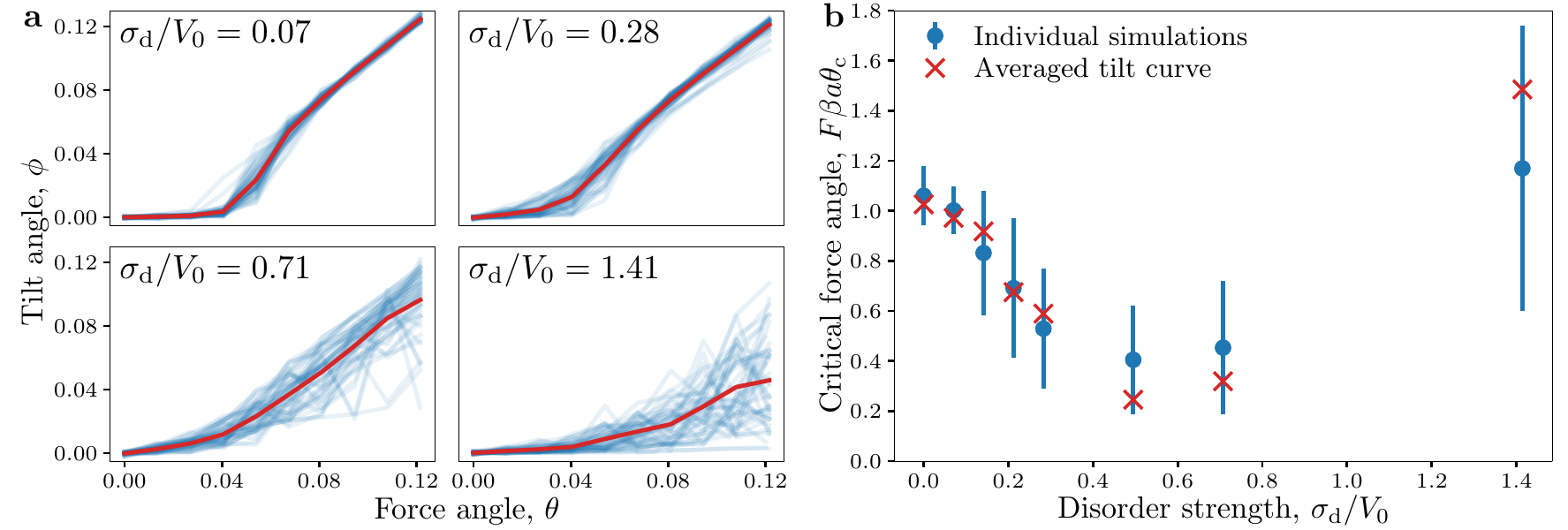}
  \caption{\subfiglabel{a}, Tilt vs. force angle measurements from
    simulations with different disorder strengths ($V_0 = 0.239$, $F=20$,
  $\beta = 1$, $N=10$). Each panel shows curves from
    50 independent random realizations of potentials (thin blue curves). The thick curve shows the average of all tilt
  values for each applied force angle. All panels have the same value ranges for
  the $\theta$ and $\phi$ axes.
  \subfiglabel{b}, Dependence of the critical force angle on disorder strength.
  Discs with error bars show the average and standard deviation of $\tcrit$
  measured from the 50 independent disorder realizations at each value of
  $\sigma_\text{d}$. Crosses show $\tcrit$ measured from the
  average tilt [thick curves in panel \subfiglabel{a}].} 
  \label{fig:disorder}
\end{figure*}

\section{Effects of quenched substrate disorder}\label{sec:disorder}
The present model of fluctuating lines under shear was originally introduced to
study the competition between non-Hermitian delocalization and \emph{Anderson}
localization~\cite{hatano1996localizationprl} of the polymers due to disorder in the substrate potential.
We now investigate the interplay of our band-insulator localization mechanism with
Anderson localization by introducing a random component to the substrate
potential $V(x)$ which is constant along the $\tau$ direction (see \appref{app:sim} for implementation
details). The disorder strength was quantified using the root mean square
amplitude $\sigmad$ of the random potential $V_\text{d}(x)$ added to the periodic
substrate,
$\sigmad \equiv \sqrt{\int_0^{Ma} \! dx\, [V_\text{d}(x)]^2/Ma}$. At each disorder
strength, multiple independent realizations of the random disorder potential were simulated; the results
are shown in \figref{fig:disorder}. While outcomes varied among
independent runs for a given disorder strength because of the finite system size
[blue curves in \subfigref{fig:disorder}{a}], we find that the mean critical force varies
nonmonotonically with disorder strength, first falling and then increasing (\subfigref{fig:disorder}{b}). A similar trend
was obeyed by critical force angles extracted from the averaged $\phi$--$\theta$
curve obtained by averaging the measured tilt angles across all independent
realizations (red curves and symbols in \figref{fig:disorder}).

These observations can be explained by considering the separate effects of disorder on the bandgap and on the localization properties of the single-particle eigenstates. Numerical studies of the lattice Hatano-Nelson model with a periodic potential~\cite{hebert2011hatanoNelson} have shown that small amounts of disorder reduced the real-valued energy gap between bands along the $\rea(\varepsilon)$ axis without affecting the extended nature of the Bloch eigenstates near the band edges.
As a result, we expect low levels of disorder to shift the delocalization transition to smaller shear values due to the reduction
of the bandgap. For a given shear value, however, all single-particle eigenstates  become localized at high-enough disorder due to Anderson localization, and the many-body fermionic ground state would also be localized even in the absence of an energy gap between unoccupied and occupied states~\cite{Basko2006}. Higher values of tilt are necessary to drive the non-Hermitian delocalization of the single-particle eigenstates at large disorder, leading to an increase in the threshold shear value. The non-monotonic behavior is consistent with a switch in the dominant localization mechanism, from band-insulator physics at low disorder to Anderson localization of single-particle eigenstates at high disorder.

The non-monotonic variation in localization with disorder strength implies that for some values of the force angle, the system undergoes two transitions as the disorder is increased. Consider a commensurate system with a force angle maintained at a value for which the polymers are localized in the absence of disorder. The system can be driven into a delocalized state by increasing the substrate disorder beyond the disorder-driven gap closure. If the disorder level is increased even further, then the polymers will eventually recover their vertical confinement due to Anderson localization of the single-particle eigenstates. This mechanism represents a non-Hermitian version of a reentrant localization transition, reminiscent of similar phenomena in Hermitian systems~\cite{Fertig1990,Roy2021,Padhan2022} and in biased Brownian motion under periodic potentials with weak disorder~\cite{reimann2008weakdisorder}.

\section{Relation to topologically quantized transport}\label{sec:adiabatic}

Since the force angle $\theta$ is shared among all polymers in the system, the transverse component of the tension can be eliminated by rotating the coordinate system so that the $\tau$ axis aligns with the tension direction. In this rotated frame, the potential energy grooves are no longer aligned with $\tau$, giving rise to a substrate potential which depends on both $x$ and $\tau$, and the corresponding quantum system becomes time-dependent. That is, the energy functional in the rotated frame is
\begin{equation}
    E = \sum_{n=1}^N \int_0^L d\tau \bigg (\frac{F}{2} (\partial_{\tau}x_n)^2 + V(x_n + \theta \tau)\bigg),
\end{equation} 
with the interaction terms omitted since they can be absorbed into the fermionic statistics. Prior work on a shear-free version of the directed polymer model with substrate potentials varying in both axes uncovered topological phenomena enabled by thermal fluctuations and noncrossing interactions~\cite{pedro2019protection}. In this section, we outline the possibility and the potential complications in realizing topologically protected chain conformations enabled by non-Hermitian spacetime-periodic potentials.

Upon performing the classical-quantum mapping as before, the quantum Hamiltonian for each fermionic particle acquires a time-dependence in the potential,
\begin{equation}\label{hamiltonian2}
    \hat H = \frac{p^2}{2m}  + V (x + ig t/m ).
\end{equation}
and the partition function of the chain satisfies the following equation~\cite{li2022slowly}:
\begin{equation}\label{floquetpartial}
    \frac {\partial \Psi(x, \tau)}{\partial \tau}  = \bigg ( \frac{1}{2 F\beta} \frac{\partial^2}{\partial x^2} -\beta V(x+\theta \tau) \bigg ) \Psi(x, \tau).
\end{equation}
The potential is periodic in space (with period $a$) and along the imaginary time axis (with period $a/\theta)$, making \eqnref{floquetpartial} a Floquet partial differential  equation~\cite{kuchment2012floquet}. The equation can also be derived by changing variables from $(x, \tau)$ to $(x', \tau') = (x + \theta \tau, \tau)$ in \eqnref{diffusion}.

At small values of $\theta$, the Hamiltonian varies slowly in time so we can use the quantum adiabatic theorem to describe the evolution of the probability density along the vertical axis. It is known that a 1D quantum system with a potential varying slowly in time exhibits a current which is quantized to multiples of the Chern number. This is called the quantized adiabatic pump or Thouless pump~\cite{thouless1983quantization}. In the polymer system, this mechanism translates to the tilt of the polymers at commensurate filling being proportional to the Chern number of the lowest band of the spacetime-periodic potential~\cite{pedro2019protection}, which equates to one for a sliding potential of the form $V(x,\tau) = V(x+\theta \tau)$~\cite{thouless1983quantization}.  Each polymer contour, on average, shifts to the right by one lattice step for each period in the $\tau$ direction, which corresponds to the the polymer profiles being tilted by an angle $\theta$ away from the vertical direction and following the potential grooves exactly. This topological tilt matches the commensurate conformations observed at low force angles in simulations.

In the rotated frame, the delocalization transition is triggered by increasing the tilt of the grooves relative to the vertical (tension) direction beyond a threshold angle. Past the transition, chains do not follow the grooves but instead align themselves closer to the vertical direction. The Chern number of the substrate potential no longer dictates the alignment of the polymers, and the transition can be interpreted as a non-Hermitian breakdown of the topological adiabatic pump. 

We have not achieved a quantitative understanding of this delocalization transition in the rotated frame, where no gap closure is apparent. The instantaneous spectrum of the time-dependent Hamiltonian in the rotated frame,
\begin{equation}
    \bigg ( \frac{1}{2 F\beta} \frac{\partial^2}{\partial x^2} -\beta V(x+\theta \tau) \bigg ) \Psi(x, \tau) = \varepsilon(\tau)\Psi(x, \tau), 
\end{equation}
can be found by replacing $x+\theta \tau \rightarrow x'$. This gives us
\begin{equation}
    \bigg ( \frac{1}{2 F\beta} \frac{\partial^2}{\partial x'^2} -\beta V(x') \bigg ) \Psi(x'-\theta\tau, \tau) = \varepsilon(\tau)\Psi(x'-\theta\tau, \tau), 
\end{equation}
so that $\Psi(x'-\theta\tau, \tau)$ is a Bloch state $e^{ik(x'-\theta\tau)}u_k(x'-\theta\tau)$ for the Hermitian Hamiltonian $\frac{1}{2 F\beta} \frac{\partial^2}{\partial x'^2} -\beta V(x')$ and $\varepsilon(\tau)$ is the corresponding real-valued energy with no dependence on $\theta$ or $\tau$. In other words, the instantaneous spectrum of the time-dependent Hamiltonian in the rotated frame, \eqnref{hamiltonian2}, does not exhibit a gap closure at any value of $g$ so an alternative mechanism for an abrupt transition at a threshold force angle is needed. One possible mechanism is a breakdown in adiabaticity due to mixing with the higher bands when the potential varies so quickly in the vertical direction that ground-state dominance no longer applies. In this limit, we must consider the Floquet spectrum of the Hamiltonian, which describes the evolution of eigenstates over full periods in the vertical direction~\cite{Bukov2015}. The breakdown of adiabaticity and corresponding delocalization could be triggered by the appearance of a Floquet exceptional point~\cite{longhi2017floquet} at a threshold force angle.

\section{Discussion}\label{sec:discussion}
In summary, we have described a new non-Hermitian delocalization transition in a statistical mechanical system of polymer chains. The transition occurs from an insulator-like localized state created via a periodic potential and noncrossing interactions, and is driven by transverse forces that generate non-Hermitian terms under a mapping to a solvable quantum Hamiltonian. We found that the delocalization of the polymer chains is caused by a gap closure in the complex non-Hermitian band structure. We derived the exact value of the critical shear in terms of the branch point structure of complex energy bands of the Hamiltonian. We also found that the critical exponent by which the shear modulus diverges is given by generic properties of the branch point. We have investigated the localization due to the combined effect of both  periodicity and disorder and uncovered preliminary evidence that while disorder favours localization at high values, it encourages \emph{de}localization at lower values. Finally, we mapped the system to a 1D non-Hermitian Thouless pump, whose breakdown triggers the delocalization transition.

Our work shows that non-Hermitian band physics~\cite{shen2018topological} and phase transitions driven by exceptional points~\cite{fruchart2021nonreciprocal}, both typically associated with driven systems, can be realized in a purely classical equilibrium setting. The directed polymer model studied here serves as a
test-bed for exploring non-Hermitian physics that is straightforward to describe and visualize, and which admits exact solutions even in the presence of thermal fluctuations and interactions. Our model and analytical framework be generalized to more complex potentials~\cite{jiang2019interplay,liu2020generalized} and interactions~\cite{Zhang2022}, as well as disorder models that could harbor new many-body localization phenomena~\cite{heussen2020manyBody}. The gauge transformation that links the spectrum of our system to the analytic continuation of band structures to complex-valued crystal momenta~\cite{kohn1959analyticProperties,heine1963surface} could provide insights into non-Hermitian delocalization in systems where localization is caused by disorder or interactions.
Higher-dimensional generalizations of our system, which generate particles with exotic statistics under the classical-quantum mapping~\cite{souslov2013polymer}, could be used to probe the non-Hermitian physics of composite fermions. 

A promising avenue for further investigation involves analyzing the system in a rotated frame which aligns the $\tau$ axis with the net force direction rather than the potential grooves. In this frame, the corresponding quantum system becomes time-dependent, opening up the possibility of realizing non-Hermitian topological phenomena~\cite{gong2018topological,shen2018topological} due to adiabatic pumping of the underlying probability distributions~\cite{pedro2019protection, li2022slowly}. Delocalization in the rotated frame could provide a manifestation of Floquet exceptional point physics~\cite{longhi2017floquet} in a classical model. The introduction of disorder, which enables unique non-Hermitian topological indices based on winding numbers~\cite{Shnerb1998,sarkar2022disorder}, provides yet another target for future studies.

While we have focused on the theoretical description of a model polymer system in our work, the transition we have uncovered could potentially be realized in a variety of fluctuating-line systems. The 1+1D statistical mechanics of vortex lines has previously been measured in type-II superconductors in a slab geometry~\cite{Bolle1999}. The substrate potential for the vortices can be controlled by patterning the superconducting properties via techniques such as focused ion beam milling~\cite{karapetrov2005direct} or masked ion irradiation~\cite{haag2014strong}. In this system, the localized state at nonzero tilt angle signifies a misalignment between the external magnetic field direction (the direction of applied tension) and the direction of the internal magnetization (carried by the vortices, which are aligned to the potential grooves)---a transverse Meissner effect~\cite{hwa1993disorder}, owed to band insulator physics rather than disorder. Other experimental candidates include artificial polymer-like fluctuating chains assembled from mesoscopic monomers such as colloids~\cite{McMullen2018,Stuij2022} and nanoparticles~\cite{Liu2020}, which can be confined to planar substrates and subjected to patterned electromagnetic or chemical forces.

\begin{acknowledgments}
  \acktext
  
\end{acknowledgments}

\appendix

\section{Simulation methods} \label{app:sim}

We implemented Langevin dynamics simulations of discretized polymer chains using
a modified version of the open-source simulation software \texttt{HOOMD-Blue}~\cite{Anderson2020},
with modifications made to enable the addition of periodic potentials of
arbitrary phase.  Each polymer is  approximated as a chain of 200 particles of mass $m$, connected by stiff
harmonic springs with equilibrium length $l_0$, implemented as a bond potential
$V_\text{bond}(r) = K(r-l_0)^2/2$
where $r$ is the distance between adjacent particles on the chain and $K$ is a
stiffness constant. The noncrossing constraint is enforced by adding a stiff contact
interaction between all pairs of particles in the system, with pair potential
$V_\text{contact}(r) = K (r-l_0)^2$ for separations $r < l_0$. For simplicity,
the same stiffness coefficient is used for both potentials. 

The tension is implemented by applying the requisite forces on the first and
last particles of each polymer chain in the desired shear angle relative to the
vertical direction. To prevent the finite-length chains from drifting
vertically, the first particle of each chain is confined to a $\tau$ coordinate
of zero with a deep and narrow harmonic potential well; the well does not
constrain the horizontal motion of the particles. The substrate potential energy
per unit length of the chain, $V(x) = V_0 \cos (2\pi x/a)$, is implemented by adding a position-dependent
potential energy of magnitude $l_0V(x)$ to each particle.

In all our simulations, we set $m=1$ and $a=1$ to set the mass and length
scales. The time scale is implicitly defined by setting $K=10000$ in simulation
units for the bond and contact stiffnesses across all simulations. We also set
$l_0 = 0.2$, so chains with 200 particles have an equilibrium length of $40a$.
The simulation box has periodic boundary conditions along the $x$ direction with
dimension $L_x = 10a$ (ten repetitions of the periodic potential) and the system
size in the $\tau$ direction is set to be much larger than the chain length. 

The equilibrium behavior of the system is simulated by using the built-in
Langevin integrator of HOOMD-Blue, which introduces random forces on each
particle that replicate the effect of a finite temperature $T$. Langevin
dynamics requires the introduction of drag forces on each particle,
$\mathbf{F}_{\text{drag},i} = - \gamma \mathbf{v}_i$ proportional to the
instantaneous velocity $\mathbf{v}_i$ of the $i$th particle. The value of the
drag coefficient affects the transient dynamics as equilibrium is approached,
but is not expected to affect the equilibrium properties. We choose a drag
coefficient $\gamma = 0.5$ for our simulations. The time step is chosen to be
0.001 in simulation units. All simulations are run for $H=10^7$ time steps or
more. To aid the evolution to equilibrium conformations, the system is
``annealed'' by starting the simulation at a temperature of $1.5 T$ and ramping
the temperature down to the desired value $T$ over the first $H/2$ time steps.
Equilibrium density profiles are then built up by sampling particle positions
during the latter $H/2$ time steps in intervals of $10^4$ time steps.

Disorder in the substrate potential is implemented by adding $n_\text{d}$ cosine potentials
with random amplitudes $\alpha_i$, wave numbers $p_i$, and phases $\phi_i$ to $V(x)$:
$$V_\text{disorder}(x) = \sum_{i=1}^{n_\text{d}} \alpha_i \cos \left(\frac{2 \pi
    p_i}{L_x} x + \phi_i\right),$$
where $\alpha_i$ are drawn from a uniform distribution chosen to generate the
desired RMS amplitude $\sigma_\text{d}$, $\phi_i$ are drawn uniformly from the
interval $[0,2\pi)$, and $p_i$ are drawn uniformly from the range $2 \leq p_i
\leq 20$.

\section{Extracting critical force angle from simulation
  data} \label{app:measurement}

\begin{figure*}
  \centering
  \includegraphics{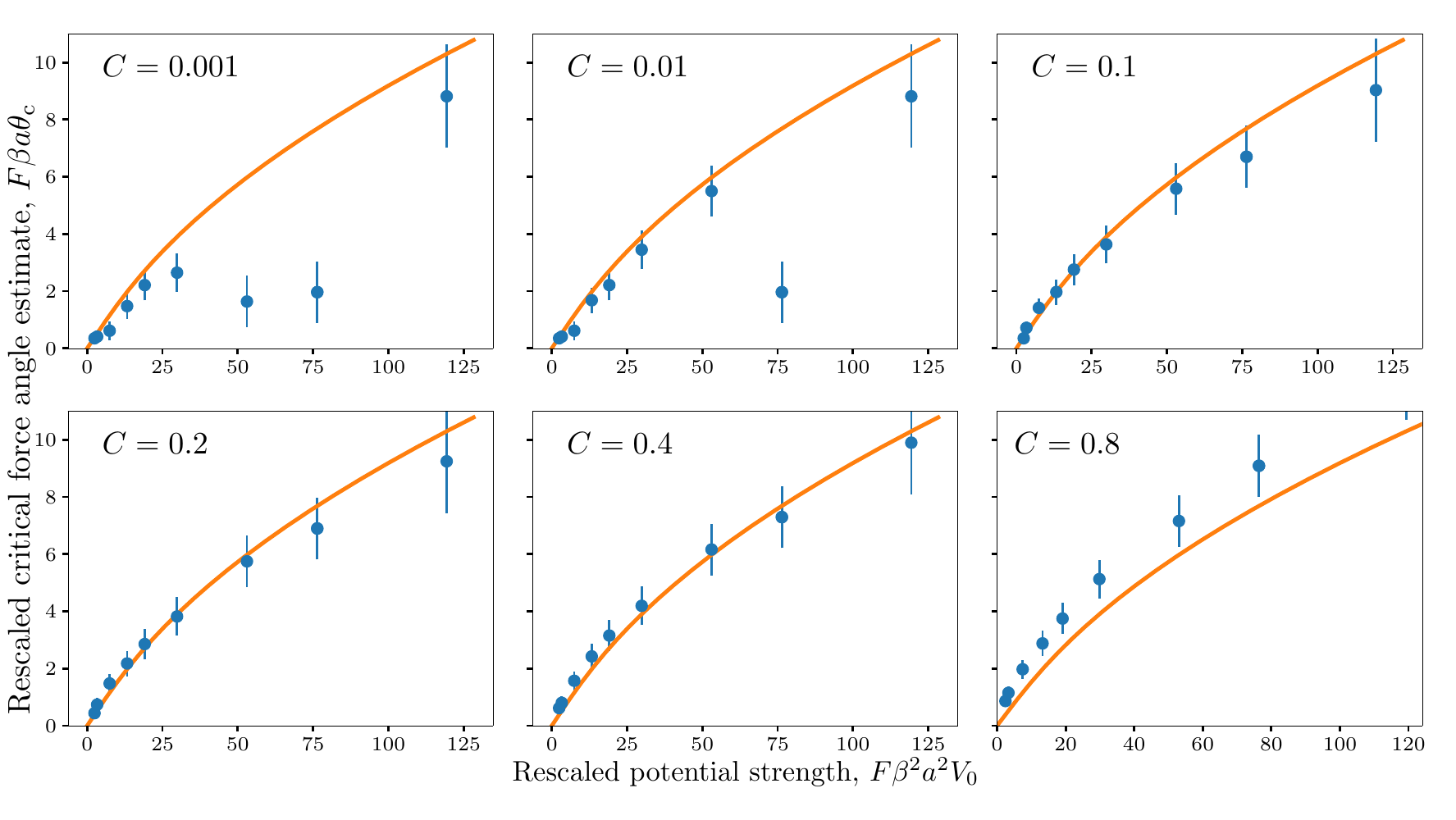}
  \caption{Critical force angle $\tcrit$ estimated as the intersection of the $\phi$-$\theta$ curve with the line $\phi = c\theta$, for different values of the numerical coefficient $c$. Estimates from simulated curves (symbols) are compared to the theoretical prediction (solid line) after rescaling as in \figref{Fig:phase}. Data shown are from the sweep with $V_0 = 0.239$ and varying $\beta$.}
  \label{fig:critslopetest}
\end{figure*}

According to our theoretical analysis, in the limit of infinitely long polymers at commensurate filling the $\phi$-$\theta$ curve should be exactly zero up to the critical value $\tcrit$, then increase with a diverging slope (as $(\theta - \tcrit)^{1/2}$) before approaching the $\phi=\theta$ line. In our finite-sized molecular dynamics simulations, the polymers acquire a slight tilt at low force angles, which abruptly increases at a finite $\theta$ value (see solid curve in \subfigref{fig:sims}{g} for the typical behavior). The small nonzero tilt at low $\theta$ arises due to bending confined to the polymer ends, whereas the steep rise (signaled by a sudden increase in the slope of the $phi$-$\theta$ relationship) is interpreted as a finite-size signature of the sharp delocalization transition in the thermodynamic limit.

To automate the estimation of the critical delocalization angle from simulations, we need a criterion to identify the abrupt rise in the polymer tilt angle curve. One option would be to set a threshold value of $\phi$ and identify $\tcrit$ as the first $\theta$ value at which the measured tilt angle is above this threshold. However, the magnitude of the tilt at low $\theta$ values depends on the system parameters such as the polymer length and potential strength, so any such threshold $\phi$ value would have to be adjusted for each simulation to accurately capture the sharp increase. Furthermore, such a criterion would not incorporate the information also present in the changing slope of the $\phi$-$\theta$ curve. Another possible approach would be to directly estimate the slope of the measured $\phi$-$\theta$ curve and apply a threshold value to the slope, but this approach is limited in precision by the large spacing between simulated $\theta$ values (restricted by the computational resources available).

Rather than imposing a threshold on the value of $\phi$ or the slope of the $\phi$-$\theta$ curve, we found that the sudden increase in tilt was reliably captured by searching for the first point of intersection between the $\phi$-$\theta$ curve and the line $\phi = C\theta$, where $C<1$ is a numerical prefactor. If $C$ is set to be large enough, the intersection point avoids the range of slow increase in tilt at low $\theta$, and correctly captures the abrupt increase in slope near the purported critical force angle. The expected large-angle behavior meanwhile restricts $C$ to be smaller than one.
If our estimation is robust, we would expect to find an intermediate range of $C$ values for which the intersection, and therefore the $\tcrit$ estimate, does not significantly change with $C$ because the two curves cross within the region of steep increase in $\phi$ with $\theta$. We indeed find that our criterion generates $\tcrit$ estimates that do not change significantly in the range $0.1 \lesssim C \lesssim 0.4$ (\figref{fig:critslopetest}). When $C$ is much smaller than 0.1, the point of intersection falls within the region of shallow slope in the $\phi$-$\theta$ for some parameter values and the resulting estimate varies strongly with $C$. At $C$ values larger than 0.4, the point of intersection falls far to the right of the region of steep increase, leading to a systematic overestimate of the critical angle. These patterns are apparent in the variations in estimated $\tcrit$ as the value of $C$ is changed in \figref{fig:critslopetest}. We use the finite spacing of the simulated $\theta$ values to quantify the uncertainty in this estimate.

In the main text, we use the curve-line intersection criterion with $C=0.1$ (the smallest value which reliably captures the abrupt increase in $\phi$ across all simulations) to extract critical angle estimates for comparison with the theoretical results in \figref{Fig:phase}. However, our conclusions would be unchanged if we used other values of $C$ within the range $0.1 \lesssim C \lesssim 0.4$, as the estimates would still agree with the theoretical prediction within the uncertainty, as shown in \figref{fig:critslopetest}.  Note that there are no fitting parameters; the theory curve is completely determined by the system parameters.

\section{Deriving the diffusion equation}\label{app:diffusion}

To obtain the differential equation whose solution provides the partition function in \eqnref{eq:pathintegral}, we consider the evolution of $\Psi$ for a small change in the $\tau$ coordinate, $\tau \to \tau + \epsilon$. Since the energy of the polymer is built up of purely local terms,  $E[ x; 0, \tau + \epsilon] = E[ x; 0, \tau] + E[ x; \tau, \tau + \epsilon]$ for any $\epsilon$. As a result, $\Psi$ obeys the useful `Markovian' recursive relation~\cite{matsen2005selfConsistent}, 
\begin{equation}
    \Psi(x_\tau, x_0, \tau + \epsilon) \sim \int \! dx_\epsilon \Psi(x_\tau, x_\tau + x_\epsilon, \epsilon)  \Psi(x_\tau + x_\epsilon, x_0, \tau).
\end{equation}

The partition function can now be evaluated iteratively ~\cite{matsen2005selfConsistent}. For small $\epsilon$ we can expand the left hand side as $ \Psi(x_\tau, x_0, \tau) + \epsilon \partial_\tau\Psi( x_\tau, x_0, \tau) + \mathcal{O}(\epsilon^2)$. We also perform the expansion,
\begin{equation}
    \Psi(x_\tau + x_\epsilon, x_0, \tau) \approx \bigg (1 + x_\epsilon \partial_x + \frac{x_\epsilon^2}{2} \partial^2_x\bigg )\Psi(x_\tau, x_0, \tau).
\end{equation}

Finally as the field does not change appreciably during the evolution by $\epsilon$ we replace $E[ x; \tau, \tau + \epsilon]$ by its mean value to get,
\begin{equation}
    \Psi(x_\tau, x_\tau + x_\epsilon, \epsilon) \sim e^{-\beta \epsilon[ \frac{F}{2} (-\frac{x_\epsilon}{\epsilon} - \theta)^2 + V(x_\tau)]}.
\end{equation}

(We change notation by omitting the initial condition $x_0$ and replacing $x_\epsilon$ by $x$.)

Plugging everything in, performing the Gaussian integral, and discarding $\mathcal O (\epsilon^2)$ terms gives the diffusion equation, \eqnref{diffusion}. Note that this connects to the assumption of the existence of a microscopic scale over which the external fields are constant and the polymer segment obeys Gaussian statistics.

\section{The full many-body system}\label{sec:manybody}

The full many-body problem is easily solved once the single-body wave functions above are known. Using Girardeau's mapping~\cite{girardeau1960mapping}, the many-body Hamiltonian, \eqnref{hamiltonian} reduces to a sum of single-body Hamiltonians provided the many-body wave function obeys the constraint, $\Psi(\bs x) = \Psi(x_1, x_2, ..., x_N) = 0$ whenever any $x_i = x_j$. This constraint is satisfied by the Slater determinant of the single-body wave functions (the Bloch waves with complex momenta).

To get real-valued solutions, we note that since the Hamiltonian, \eqnref{hamiltonian}, is real, its normalized eigenstates $\Psi_n(\bs x)$ are either real or come in complex conjugate pairs. Thus we just need to ensure that the Slater determinant has either real eigenstates or pairs of complex conjugate ones.

For a filled band it is useful to define the solution in terms of Wannier functions rather than Bloch waves. The descriptions are equivalent since the determinant of a matrix of solutions is invariant on multiplication with a Unitary matrix. The Slater determinant in terms of the single-body Wannier functions, $\Phi_X(x)$, is
\begin{equation}
    S(\bs x) = \frac{1}{\sqrt{N!}} \sum_{\sigma \in S_N} \text{sgn}(\sigma) \prod_{i=1}^N \Phi_{X_i}(\sigma(x_i)),
\end{equation}
where the Wannier functions,  $\Phi_X(x)$, are related to the Bloch functions,  $\Psi_k(x)$, by 
\begin{equation}
    \Phi_X(x) = \frac{1}{\sqrt N} \sum_k e^{-ikX} \Psi_k(x); X \in \{a, 2a, \dots, Ma \}.
\end{equation}

While the above defined functions are not unique (due to the freedom in choice of global phase for the Bloch functions) a unique set of real-valued Wannier functions can always be found~\cite{kohn1959analyticProperties}.

Finally, since the polymers are distinguishable we restrict the domain of the constructed wave-function:
\begin{equation}\label{polymermany}
    \Psi(\bs x) = \begin{cases}
    \sqrt{N!}S(\bs x) &\mbox{if } \bs x \in \mathcal{R}_0 \\
0 & \mbox{elsewhere, } \end{cases}
\end{equation}
where $\mathcal{R}_0$ is defined by the inequalities, $x_1 < x_2 < ... < x_N \text{ (mod } Ma)$, 
and is the physical region allowed in our non-crossing problem~\cite{gennes1968soluble}. This redefining does not interfere with the wave function being an eigenstate since it still satisfies the eigenvalue equation and is still continuous. (The derivative of the wave function is allowed to be discontinuous because of the singular terms, $\delta(x_i - x_j)$, in the Hamiltonian.)

Note that while Girardeau's fermion-to-boson mapping requires the determinant $S(\bs x)$ be multiplied with an anti-symmetry factor, $A(\bs x) \in \{\pm 1\}$, to render it symmetric~\cite{girardeau1960mapping}, this is not required since our wave function is non-zero only in $\mathcal{R}_0$.

\section{Computation of band-structure and the critical angle}\label{app:computation}

The band structure, $\varepsilon_n(k)$ at complex $k$ can be computed using any electronic-structure calculations software such as Ref.~\cite{giannozzi2009quantum} as used in Ref.~\cite{smogunov2004conductance}.

In our case, we used the fact that exact solutions for the cosine potential, $V(x) = V_0\cos(2\pi x/a)$ are known in terms of the Mathieu functions~\cite{kovacic2018Mathieu}, which are implemented with high precision in computational software such as \texttt{Mathematica}. To generate \figref{Fig:bandStructure}, \figref{fig:phitheta}, and \figref{fig:riemann}, we computed the Floquet exponent (also called the Mathieu characteristic exponent for the specific case of the cosine potential) for different complex values of the energy eigenvalue $\varepsilon$. The imaginary part of the Floquet exponent is the same as $\ima k$ up to numerical factors. After computing a fine mesh of Floquet exponents for a range of energy eigenvalues, we interpolate the values to generate smooth sheets/curves. For \figref{Fig:phase}, we used the fact that $\varepsilon_n(k)$ is real for $k$ in the line segment joining $\frac{\pi}{a} - i\frac{\mu}{a}$ to $\frac{\pi}{a} + i\frac{\mu}{a}$~\cite{heine1963surface}. To find $\mu$ for any fixed value of potential strength, $V_0$, we then had to increase the value of $\ima k$ (with $\rea k$ fixed to $\pi/a$) until the corresponding value of $\varepsilon$ became complex-valued. 

\section{Derivation of polymer tilt}

\subsection{Tilt from energy eigenvalues} \label{app:tilt}

In Refs.~\onlinecite{hatano1996localizationprl,hatano1997vortexPinning}, the tilt angle of a polymer in a single-particle eigenstate $\Psi_m$ of the Hamiltonian $\Hhn$ was shown to be related to the dependence of the energy eigenvalue on the force angle: $$\phi = -\frac{\partial \varepsilon_m}{\partial g} = -\frac{1}{F}\frac{\partial \varepsilon_m}{\partial \theta}.$$ The expression was derived by defining a current operator in terms of a derivative of the Hamiltonian with respect to the vector potential strength $g$. Here, we provide an alternative derivation of this expression using classical probability currents under periodic boundary conditions, and also show that it describes the average tilt of the many-body system.

We begin by evaluating the change in the probability density function along the $\tau$ direction:
\begin{align*}
 &\frac{\partial p(x,\tau)}{\partial \tau} \equiv \partial_\tau p(x,\tau)  = \frac{1}{Z}\left[ \Psit \partial_\tau \Psi + \Psi \partial_\tau \Psit \right] \\
  &=\frac{1}{Z}\left[ \frac{1}{2\beta F}\left(\Psit \partial_x^2 \Psi - \Psi \partial_x^2 \Psit\right) -\theta \left(\Psit \partial_x \Psi + \Psi \partial_x \Psit\right) \right].  
\end{align*}
This change in density generates a local probability current $j(x,t)$ along the $x$ direction through the continuity equation
\begin{align*}
  \partial_\tau p + \partial_x j &= 0 \\
  \Rightarrow j(x,\tau) &= \frac{1}{Z} \left[-\frac{1}{2\beta F}\left(\Psit \partial_x \Psi - \Psi \partial_x \Psit\right) +\theta \left(\Psit  \Psi\right) \right].
\end{align*}
The integrated probability current provides the rate of change in the average polymer position along the $\tau$ direction, which is the local tilt angle: $\partial_\tau \langle x \rangle = \int \! dx\, x \partial_\tau p = -  \int \! dx\, x \partial_x j =  \int \! dx \, j$, where the last step involves integration by parts on the periodic domain $0 \leq x < L_x$. Therefore, we obtain
\begin{align}
  \partial_\tau \langle x \rangle &= \int_0^{L_x} \! dx \, j(x,\tau)\nonumber \\
  &= -\frac{1}{2\beta F Z}\int_0^{L_x} \! dx  \left(\Psit \partial_x \Psi - \Psi \partial_x \Psit \right) +\frac{\theta}{Z}\int_0^{L_x} \! dx\,  \Psit \Psi\nonumber \\
  &= -\frac{1}{\beta F Z}\int_0^{L_x} \! dx  \left(\Psit \partial_x \Psi\right)  +\theta\label{thetainter1},
\end{align}
where the last step again involves integration by parts. To understand the origin of the constant term, consider a single polymer wandering across a featureless substrate $V(x)= 0$ at a nonzero force angle with free ends. The aggregated density profile is a constant over all space so $\partial_x \Psi = 0$, yet the polymer aligns to the force direction on average so the tilt equates to $\theta$.

To express the above equation in terms of the eigenvalues we first note that,
\begin{align}
    \Hhn &=  \frac{(p + ig)^2}{2m} + V(x) \nonumber\\
    &= -\frac{1}{2\beta^2F}\partial_x^2 + \frac{\theta}{\beta}\partial_x - \frac{F\theta^2}{2} + V(x),
\end{align}
such that,
\begin{equation}
    \partial_\theta \Hhn = \frac{1}{\beta}\partial_x -F\theta .
\end{equation}

Now we expand the equation $\partial_\theta (\mathcal{H}\Psi_n) = \partial_\theta(\varepsilon_n \Psi_n)$, multiply both sides by $\Psit_m(x)$ and integrate,
\begin{equation}
    \int \! dx \,(\Psit_m(\partial_\theta\mathcal{H})\Psi_n + (\varepsilon_m- \varepsilon_n) \Psit_m \partial_\theta \Psi_n) = \delta_{mn}\partial_\theta \varepsilon_m.
\end{equation}
We used $\mathcal{H}^\dagger \Psit_m^* = \varepsilon_m^* \Psit_m^*$ to simplify $\Psit_m \mathcal{H} = \varepsilon_m \Psit_m.$ If $m=n$ (we will be using ground-state dominance), we get,
\begin{equation}\label{partialE}
    \partial_\theta \varepsilon_m = \int \! dx \,\Psit_m(\partial_\theta\mathcal{H})\Psi_m = \frac{1}{\beta}\int \! dx \,\Psit_m\partial_x\Psi_m - F\theta.
\end{equation}

\begin{figure*}
  \centering
  \includegraphics{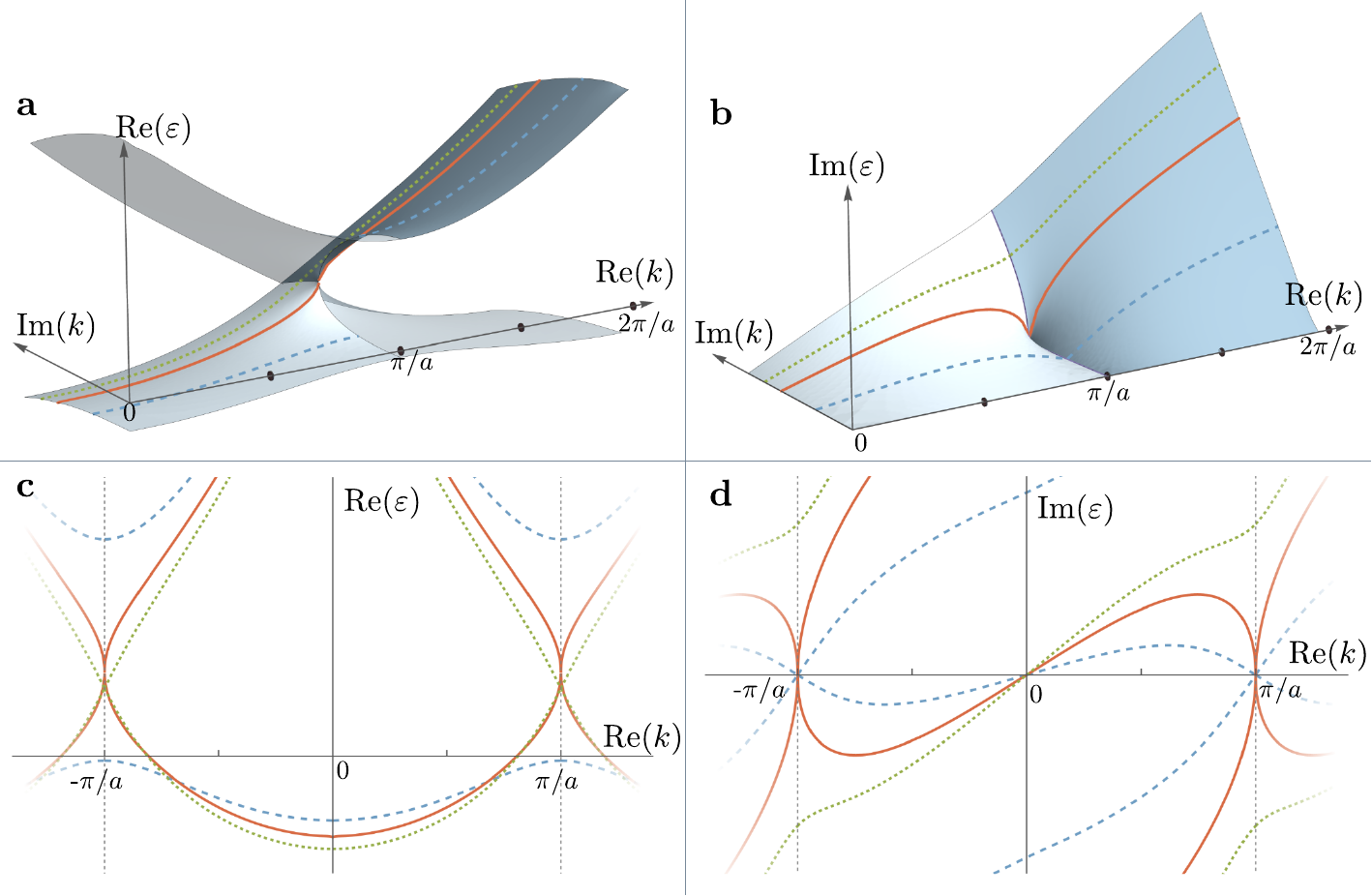}
  \caption{The complex band-structure $\varepsilon(k)$ of the Hamiltonian, $\frac{p^2}{2m} + V(x)$ with $V(x) = \cos(x)$ as a function of complex $k$. Panels \subfiglabel{a} and \subfiglabel{b} show the real and imaginary parts, respectively, of the lowest two bands (Riemann sheets ordered by the real part of the energy) of $\varepsilon(k)$. The lowest band is plotted as a solid surface in the interval $0 < \rea k < \pi/a$ and the next band is plotted in the interval $\pi/a < \rea k < 2 \pi /a$. In panel \subfiglabel{a}, the other band in each region is shown as a translucent surface. Only the region with $\rea k \geq 0$ and $\ima k \geq 0$ is shown since on changing the sign of either $\rea k$ or of $\ima k$, $\varepsilon(k)$ turns into $\varepsilon(k)^*$~\cite{kohn1959analyticProperties}. The three contours are at $\ima k = 0.5\mu/a$ (blue, dashed), $\ima k = \mu/a$ (red, solid) which corresponds to the critical shear, and $\ima k = 1.25\mu/a$ (green, dotted). Panels \subfiglabel{c} and \subfiglabel{d} show these contours in the full Brillouin zone, illustrating that $\varepsilon(k)$ is periodic upon advancing $\rea(k)$ by $2\pi/a$.}
  \label{fig:riemann}
\end{figure*}

Given specific boundary conditions at the polymer ends, we can write the tilt of an arbitrary polymer state using the spectral expansion introduced in the main text; this would give a local tilt angle dependent on the vertical coordinate $\tau$.  In the interior of a long polymer, however, ground-state dominance dictates that $(\Psi,\Psit) \to (\Psi_0,\Psit_0)$ and, on substituting \eqnref{partialE} in \eqnref{thetainter1}, the local tilt angle becomes a constant,
\begin{equation}
  \label{eq:tilt-groundstate}
\partial_\tau \langle x \rangle = -\frac{1}{F} \frac{\partial \varepsilon_0}{\partial \theta} \to \phi.  
\end{equation}
We equate this to the measured tilt angle, $\phi$, in the simulations.

For a system of many polymers, we now have $N$ probability currents $j_n$ and the continuity equation is transformed to $$\partial_\tau p(\bs x; \tau) + \sum_{i=1}^{N} \partial_{x_i} j_i(\bs x; \tau) = 0.$$ The local tilt angle associated with the $n$th polymer is obtained by integrating the corresponding current:
\begin{align*}
  \partial_\tau \langle x_n \rangle &= -\int_0^{L_x} \! d\bs x \,x_n \sum_i \partial_{x_i} j_i(\bs x;\tau) \\
  &= \int_0^{L_x} \! d\bs x \, j_n(\bs x;\tau) \\
  &= -\frac{1}{\beta F \bs Z}\int_0^{L_x} \! d\bs x  \left(\Psit \partial_{x_n} \Psi\right)  +\theta,
\end{align*}
where $\bs Z = \int \! d\bs x\, (\Psit\Psi)$. Since all the polymers are statistically identical, the equilibrium tilt of an individual polymer is the same as the average tilt of all the polymers, which decomposes into a sum of single-particle terms under the mapping we use for noncrossing polymers:
\begin{align}
  \overline{\partial_\tau \langle x_n \rangle} &= \frac{1}{N} \sum_n \left[ -\frac{1}{\beta F \bs Z}\int_0^{L_x} \! d\bs x  \left(\Psit \sum_n \partial_{x_n} \Psi\right)  +\theta\right] \nonumber \\
  &= \frac{1}{NF\bs Z}\int_0^{L_x} \! d\bs x \, \Psit \left( \partial_\theta \sum_n \mathcal{H}_n(g) \right) \Psi.
\end{align}
Under ground-state dominance the average tilt angle again depends solely on the many-body energy eigenstate with the lowest real part and we obtain \eqnref{eq:tilt-manybody-groundstate} of the main text:
$$\phi = -\frac{1}{NF} \frac{\partial \bs \varepsilon_0}{\partial \theta}.$$

\subsection{Using the analytical properties of the complex band structure}\label{app:tilt-analytical}

Eqn.~\eqref{eq:tilt-final} of the main text establishes that the tilt can be simplified to,
  $$\phi =\frac{\beta a}{\pi}\ima\left(\varepsilon_0\left(\frac{\pi}{a} + i \beta F \theta\right)\right).$$

The multivalued function $\varepsilon(k)$ is also periodic upon advancing $\rea(k)$ by $2\pi/a$~\cite{kohn1959analyticProperties}. At constant $\ima(k) =\beta F \theta$ with $\theta < \tcrit$, the Riemann sheet $\varepsilon_0(k)$ is well-separated from its adjacent sheets (which form higher bands), and is itself periodic with the same periodicity as $\varepsilon(k)$ (see \figref{fig:riemann}\subfiglabel{a}). This implies that $\varepsilon_0\left(-\frac{\pi}{a} + i \beta F \theta\right) = \varepsilon_0\left(\frac{\pi}{a} + i \beta F \theta\right)$, and the average tilt (\eqnref{eq:tilt-complex}) evaluates to zero. 

By contrast, when $\ima(k) > \mu_0/a = \beta F \tcrit$, i.e., for force angles above the critical angle, the real part of the lowest band touches the next band at $\rea(k) = \pm \pi/a$ (see \figref{fig:riemann}). Now, although the multivalued function $\varepsilon(k)$ is still periodic along the $\rea(k)$ axis, the sheet $\varepsilon_0(k)$ with the lowest real part of the energy is no longer periodic. Upon smoothly following $\varepsilon_0$ from one endpoint of the contour to the other, the real component returns to its starting value, whereas the imaginary component is nonzero at either endpoint. This feature is apparent in the shapes of the dashed lines in \figref{Fig:bandStructure} above the critical contour: the oval corresponding to the lowest band opens up when it merges with the next band, but remains symmetric about the $\rea(\varepsilon)$ axis. Therefore, the average tilt becomes nonzero when $\theta > \tcrit$, since the closed oval corresponding to the lowest band `opens up' by merging with the next band and $\varepsilon_0\left(\frac{\pi}{a} + i \beta F \theta\right)$ becomes complex-valued.

We can deduce additional features of the $\phi$--$\theta$ curve near $\tcrit$ from known analytic properties of $\varepsilon(k)$. Near the branch point $k_0 = \pi/a + i F\beta \tcrit$, the energy behaves as $\varepsilon_0(k) = \varepsilon_0(k_0) + A e^{i\frac{\pi}{4}}(k-k_0)^{1/2}$ to lowest order in $(k-k_0)$, where $A$ is a nonzero real constant~\cite{kohn1959analyticProperties,heine1963surface}. Since $\varepsilon_0(k_0)$ is real, the imaginary part of $\varepsilon_0(\frac{\pi}{a}+i \beta F\theta)$ has the following behavior near $\tcrit$:
\begin{equation}
  \ima\left(\varepsilon_0\left(\frac{\pi}{a} + i \beta F \theta\right)\right) =
  \begin{cases}
    0 & \theta < \tcrit \\
    \frac{A}{\sqrt 2} \sqrt{F\beta a (\theta-\tcrit)} & \theta > \tcrit
  \end{cases}
\end{equation}
Immediately after the transition, therefore, the tilt is expected to grow as $\phi \sim (\theta - \tcrit)^{1/2}$ in the limit of large system sizes. 

Finally, for very large values of the force angle and hence of $\ima(k)$, the momentum term dominates the Hamiltonian and we can ignore the potential term. The eigenfunctions of $\hat{H}$ are of the form $e^{ikx}$ with energy-momentum relation $\varepsilon(k) = \frac{k^2}{2\beta^2 F}$. For $\theta \gg \tcrit$, we therefore expect
\begin{align}
  \ima\left(\varepsilon_0\left(\frac{\pi}{a} + i \beta F \theta\right)\right) &= \frac{1}{2\beta^2 F}\ima \left(\left(\frac{\pi}{a} + i \beta F \theta\right)^2\right)\nonumber \\
  &= \frac{\pi}{\beta a} \theta\label{eq:largetheta},
\end{align}
from which we obtain $\phi = \theta$ from \eqnref{eq:tilt-final}.

The calculations can be easily generalized to arbitrary fillings too. For $N$ polymers and $M$ potential wells, where $N$ is odd, we get,
\begin{align} 
  \phi &= -\frac{1}{NF} \frac{\partial{\emb}_0}{\partial{\theta}} = -\frac{M\beta a}{N2\pi} \int_{-\frac{\pi(N-1)}{Ma}}^{\frac{\pi(N-1)}{Ma}} \! dk \, \frac{\partial \rea(\varepsilon_0(k))}{\partial \ima(k)}\nonumber,\\
  &=\frac{M}{N}\frac{\beta a}{\pi}\ima\left(\varepsilon_0\left(\frac{\pi(N-1)}{Ma} + i \beta F \theta\right)\right).
\end{align}
This corresponds to the path along increasing values of $\ima k$ at constant $\rea k = \frac{\pi(N-1)}{Ma}$ in \figref{fig:riemann}\subfiglabel{b}. When $N$ is even, the many-body ground-state eigenvalues and eigenstates come in complex-conjugate pairs. In that case, the generalization of the calculation requires forming appropriate superpositions of the two complex-conjugate states, in terms of their real and imaginary parts, to keep the probability density, $p(\bs x,\tau)$, real.

The calculations in this Appendix rely on generic features of the complex band structures of one-dimensional periodic potentials as detailed in Refs.~\onlinecite{kohn1959analyticProperties}, \onlinecite{heine1963surface} and \onlinecite{prodan2006analytical}. We can verify these general predictions for the specific (cosine) potential we have used in our study, for which we can compute the complex-valued energy bands following \appref{app:computation}. Figure~\ref{fig:riemann} shows the three distinct behaviors for \eqnref{eq:tilt-final} as the force angle is increased.

\end{document}